\documentclass[%
 aip, amsmath,amssymb, reprint, floatfix]{revtex4-1}
\usepackage{graphicx}
\usepackage{dcolumn}
\usepackage{bm}

\usepackage[utf8]{inputenc}
\usepackage{xcolor}
\usepackage{amssymb}
\usepackage{backnaur}
\usepackage{algorithm}
\usepackage{algpseudocode}
\usepackage{cancel}
\usepackage{url}
\usepackage{dcolumn}
\usepackage[normalem]{ulem}
\usepackage{times}
\usepackage{mathptmx}
\usepackage{multirow}
\usepackage{braket}

\usepackage[english]{babel}
\usepackage{xcolor}
\colorlet{RED}{red}
\colorlet{BLUE}{blue}
\usepackage{dcolumn}
\usepackage{bm}
\usepackage[version=4]{mhchem} 
\usepackage{acronym}
\usepackage{subfig}
\usepackage{adjustbox}
\usepackage{float}
\usepackage{tikz}
\usepackage{microtype} 
\usepackage{algorithm}
\usepackage{algpseudocode}
\usepackage{graphicx}
\usepackage{caption}
\usepackage{subcaption}
\usepackage{dcolumn}
\usepackage{bm}
\usepackage{color} 
\usepackage{epsfig}
\usepackage{bbold}
\usepackage{url}
\usepackage{soul}
\usepackage{lineno}
\usepackage[T1]{fontenc}

\usepackage{listings}
\usepackage{color}

\definecolor{dkgreen}{rgb}{0,0.6,0}
\definecolor{gray}{rgb}{0.5,0.5,0.5}
\definecolor{mauve}{rgb}{0.58,0,0.82}
\DeclareRobustCommand{\textblack}{\textcolor{black}}

\newcommand{\inlinecode}[1]{\colorbox{lightgray}{\small{\lstinline$#1$}}}
\newcommand{\code}[1]{{\ifmmode{\texttt{#1}}\else$\texttt{#1}$\fi}}
\newcommand{\boldit}[1]{{\ifmmode{\textit{\textbf{#1}}}\else$\textit{\textbf{#1}}$\fi}}
\newcommand{\response}[1]{\textcolor{black}{#1}}
\newcommand{\finalresponse}[1]{\textcolor{black}{#1}}

\newcommand{\finalchanges}[1]{\textcolor{black}{#1}}

\usepackage[colorlinks=true,citecolor={blue},urlcolor={black},linkcolor={blue}]{hyperref}

\usetikzlibrary{calc,shapes.geometric,decorations.pathmorphing,patterns}

\begin{document}



\title{TAMM: Tensor algebra for many-body methods}

\author{Erdal Mutlu}
\email{erdal.mutlu@pnnl.gov}
\affiliation{Advanced Computing, Mathematics, and Data Division, Pacific Northwest National Laboratory, Richland, Washington 99354, USA}

\author{Ajay Panyala}
\email{ajay.panyala@pnnl.gov}
\affiliation{Advanced Computing, Mathematics, and Data Division, Pacific Northwest National Laboratory, Richland, Washington 99354, USA}

\author{Nitin Gawande}
\affiliation{%
Intel Corporation, Richland, Washington 99354 USA
}

\author{Abhishek Bagusetty}
\affiliation{Argonne Leadership Computing Facility, Argonne National Laboratory, Argonne, Illinois 60439, USA}

\author{Jeffrey Glabe}
\affiliation{Advanced Computing, Mathematics, and Data Division, Pacific Northwest National Laboratory, Richland, Washington 99354, USA}

\author{Jinsung Kim}
\affiliation{%
School of Computer Science and Engineering, Chung-Ang University, Seoul 06974, South Korea
}

\author{Karol Kowalski}
\affiliation{Physical Sciences Division, Pacific Northwest National Laboratory, Richland, Washington 99354, USA}
\author{Nicholas P. Bauman}
\affiliation{Physical Sciences Division, Pacific Northwest National Laboratory, Richland, Washington 99354, USA}

\author{Bo Peng}
\affiliation{Physical Sciences Division, Pacific Northwest National Laboratory, Richland, Washington 99354, USA}

\author{Himadri Pathak}
\affiliation{Advanced Computing, Mathematics, and Data Division, Pacific Northwest National Laboratory, Richland, Washington 99354, USA}

\author{Jiri Brabec}
\affiliation{
J. Heyrovsk\'{y} Institute of Physical Chemistry, Academy of Sciences of the Czech Republic,  182 23 Prague 8, Czech Republic}

\author{Sriram Krishnamoorthy}
\email{sriram.krishnamoorthy@gmail.com}
\affiliation{Google Inc., Mountain View, California 94043, USA}

\begin{abstract}
Tensor algebra operations such as contractions in computational chemistry consume a significant fraction of the computing time on large-scale computing platforms. The widespread use of tensor contractions between large multi-dimensional tensors in describing electronic structure theory has motivated the development of multiple tensor algebra frameworks targeting heterogeneous computing platforms. In this paper, we present Tensor Algebra for Many-body Methods (TAMM), a framework for productive and performance-portable development of scalable computational chemistry methods. TAMM decouples the specification of the computation \textblack{from} the execution of these operations on available high-performance computing systems. With this design choice, the scientific application developers (domain scientists) can focus on the algorithmic requirements using the tensor algebra interface provided by TAMM, whereas high-performance computing developers can direct their attention to various optimizations on the underlying constructs, such as efficient data distribution, optimized scheduling algorithms, and efficient use of intra-node resources (e.g., graphics processing units). The modular structure of TAMM allows it to support different hardware architectures and incorporate new algorithmic advances. We describe the TAMM framework and our approach to the sustainable development of scalable ground- and excited-state electronic structure methods. We present case studies highlighting the ease of use, including the performance and productivity gains compared to other frameworks.
\end{abstract}

\maketitle

\renewenvironment{bnf}%
{\csname align\endcsname}%
{\csname endalign\endcsname\ignorespacesafterend}



\renewenvironment{bnf*}%
{\csname align*\endcsname}%
{\csname endalign*\endcsname\ignorespacesafterend}

\renewenvironment{bnf}%
{\csname align\endcsname}%
{\csname endalign\endcsname\ignorespacesafterend}

\newcommand{\bnftsfont}[1]{\texttt{#1}}
\renewcommand{\bnfpo}{::=}
\newcommand{\bnfalt}{&\bnfor}
\renewcommand{\bnfts}[1]{\;\textnormal{\bnftsfont{#1}}\;}
\renewcommand{\bnfprod}[2]{\bnfpn{#1} \; & \bnfpo \; #2}

\section{INTRODUCTION}

Enabling highly-scalable computational environments that abstract and automate the development of complex tensor algebra operations is critical in order to advance computational chemistry towards more complex and accurate formulations capable of taking advantage of emerging exascale computing architectures and also serve as a foundation for a sustainable and portable electronic structure software stack.
In particular, tensor contractions (TCs) are a universal language  used in many areas of quantum mechanics to encode equations describing collective phenomena in many-body quantum systems encountered in quantum field theory, quantum hydrodynamics, nuclear structure theory, material sciences, and quantum chemistry. Typically, contractions between multi-dimensional tensors stem from the discretization procedures \textblack{used} to represent the Schr\"odinger equation in a finite-dimensional algebraic form. An important area where tensor contractions play a crucial role is electronic structure theory, where tensors describe basic interactions and parameters defining wave function expansions.

One of the most critical applications of TCs is the coupled-cluster (CC) formalism.  
\cite{coester58_421,coester60_477,cizek66_4256,purvis82_1910,paldus72_50,paldus07,crawford2000introduction,bartlett_rmp}
In the CC theory, the form of the complex  TCs used to represent non-linear equations describing the correlated behavior of electrons in molecular systems also reflects the fundamental feature of the CC formalism referred to as the size-extensivity or proper scaling of the energy with the number of particles. For this reason, the CC formalism is one of the most accurate computational models used these days in computational chemistry. 

The CC theory has assumed a central role in high-accuracy computational chemistry and still attracts much attention in theoretical developments and numerical implementations. In this effort,  high-performance computing (HPC) and the possibility of performing TCs in parallel play an essential role in addressing steep polynomial scaling as a function of system size and extending CC formalism to realistic chemical problems described by thousands of basis set functions. To understand the scale of the problem, canonical CC formulations such as the ubiquitous CCSD(T) approach\cite{ccsd_t} (CC with iterative single and double excitations with non-iterative corrections due to triple excitations) 
involve contractions between four-dimensional tensors where ranges of thousands of entries can define each dimension. In order to alleviate efficient CC calculations on parallel platforms, several specialized tensor libraries have been developed over the last decade.

In the last few decades, significant effort has been expended toward enabling CC simulations for very large chemical systems to extend the applicability of the CC formalism further. In the reduced-scaling formulations, commonly referred to as the local CC methods,\cite{hampel1996local,schutz2000low,schutz2000local,neese2009efficient,riplinger2013efficient,dlpno-sparse,dlpno-sparse2,pavosevic2016,saitow2017new} one takes advantage of the local character of correlation effects to effectively reduce the number of parameters and the overall cost of CC iterations, allowing for calculations of systems described by 10,000-40,000 basis set functions.
%
%
This paper discusses a new tensor library, Tensor Algebra for Many-body Methods (TAMM), which provides a flexible environment for expressing and implementing TCs for canonical and local CC approximations. The development of functionalities for the ground-state formulations \textblack{[]}canonical CCSD and CCSD(T) methods and local CC formulations\textblack{]} using TAMM is supported by the NWChemEx project,~\footnote{https://nwchemex-project.github.io/NWChemEx/} whereas excited-state formulations \textblack{[}CC Green's function \textblack{(CCGF)}, Equation-of-Motion \textblack{(EOM)} CC, and {\color{black}Time-dependent CC (TDCC)}] \textblack{are supported} by the SPEC project.~\footnote{https://spec.labworks.org/home}

As HPC systems continue to evolve to include different types of accelerators, diverse memory hierarchy configurations, and varying intra- and inter-node topologies, there is a need to decouple the development of electronic structure methods from their optimizations for various platforms. TAMM enables the compact specification of various electronic structure methods that heavily rely on tensor operations while allowing independent yet incremental development of optimization strategies to map these methods to current and emerging heterogeneous platforms. 

\finalresponse{The rest of the paper is organized as follows\textblack{:} Section~\ref{sec:related_work} briefly discusses other tensor algebra frameworks used for developing computational chemistry applications. Section~\ref{sec:tamm_framework} describes the details of our tensor algebra interface and \textblack{the} underlying constructs that are used to efficiently distribute tensor data \textblack{and} schedule and execute tensor operations on modern HPC platforms. Later in \ref{sec:feature_compare_other_tensor_libs}, we provide a feature comparison with other distributed tensor algebra frameworks. 
Section~\ref{sec:case_studies} showcases multiple CC methods implemented using TAMM and  performance results obtained using the OLCF Summit Supercomputer.~\cite{olcfsummit} Section~\ref{sec:performance_comparison} demonstrates the performance of TAMM in comparison to other distributed tensor algebra \textblack{frameworks}.} 

\section{\textblack{REVIEW OF EXISTING INFRASTRUCTURE}}
\label{sec:related_work}

Tensor-based scientific applications have been widely used in different domains, from scientific simulation to more recent machine learning\textblack{(ML)}-based scientific applications. Over the years, program synthesis and code generation have become the go-to solution for such applications. The Tensor Contraction Engine (TCE),\cite{hirata_tce} which is used in the NWChem computational chemistry software package,~\cite{nwchem} has been one such successful solution for automatically generating parallel code for various molecular orbital (MO) basis CC methods in Fortran 77 .
In later work, the TCE project~\cite{tce_05} added support for optimizations on tensor expression factorization, optimized code generation for various hardware, and space time trade-offs in order to improve and also implement more complex CC methods.

In a separate effort, the FLAME project~\citep{flame_02} provided formal description support for describing linear algebra operations over matrices with support \textblack{for} the optimized implementation of these kernels for distributed memory systems. Later, various studies over-optimizing tensor algebra~\cite{summa_16,symmetric_14,elemental_13} have been proposed using the FLAME framework.

The Cyclops Tensor Framework (CTF)~\cite{ctf_14} was developed aiming \textblack{for a} more efficient kernel implementation for tensor operations using concurrency. The framework focused on reducing the required communication in parallel executions of CC-based methods by using a \textblack{dense triangular} tensor representation.  Solomonik and Hoefler extended CTF for general sparse computations.\cite{sparse_ctf_15} Manzer {\it et al.} demonstrated the benefits of exploiting block sparsity in electronic structure calculations.\cite{manzer-jctc2017} Neither approach includes notational support for the general block-structured nature of the sparsity that naturally occurs in electronic structure methods. 

Epifanovsky {\it et al.}\cite{libtensor_13} developed an object-oriented C++ library called \texttt{libtensor} for efficient execution of post-Hartree Fock electronic structure methods using a blocked representation of large-size dense tensors. In later work,~\cite{libtensor_opt} they optimized various operations \textblack{using} efficient memory management techniques that are thread-friendly and NUMA-aware.  \finalresponse{Unlike TAMM, libtensor does not offer a distributed infrastructure for tensor algebra operations and is restricted to shared memory systems.}

Orca is a general quantum chemistry program involving implementations of reduced-scaling methods. In order to achieve the speed up, various approximations are employed, for example, density fitting, RIJ-COSX, or \textblack{a} local approach for NEVPT2 or CC methods. The C++ code employs \finalchanges{message passing interface (MPI)}-based parallelization schemes, and recently, in a pilot study, they implemented a scheme for generating 3- and 4-index integrals via accelerators.\cite{orca_2020} \finalresponse{Orca provides a wide range of computational chemistry methods, whereas TAMM provides a general distributed tensor algebra framework to enable the productive development of scalable computational chemistry applications.}

\textblack{The} ExaTensor~\cite{exatensor} library uses a domain-specific virtual \textblack{processor} concept \textblack{to allow} performance portable programming and execution of tensor operations on modern \textblack{graphics processing unit} architectures. While the library is mostly focused on general GPU computation and the portability of such operations to different systems, they demonstrated \textblack{its} effectiveness on numerical tensor algebra workloads that mainly focus on quantum many-body computations on HPC platforms. \finalresponse{The main focus of the ExaTensor library is to utilize GPUs for efficient tensor contractions. TAMM, on the other hand, provides a large set of supported tensor operations and utility routines. TAMM also provides special constructs allowing users to add new operations on tensors that are needed for the development of complete computational chemistry applications.}

The TiledArray framework~\cite{tiled_array} is actively being developed for scalable tensor operations \textblack{that support} various computational chemistry methods.~\cite{ta_16} It makes use of the Multi \textblack{r}esolution Adaptive Numerical Environment for Scientific Simulation (MADNESS) parallel runtime\textblack{, which} ~\cite{madness_16} that employs a high-level software environment for increasing both programmer productivity and code scalability in massively parallel systems. TiledArray employs a hierarchical view of sparsity coupled with explicit user-written loop nests to perform specialized operations over sparse tensors. \finalresponse{TiledArray has several similarities with TAMM while also differing in several key aspects, which are explained later in Section \ref{sec:feature_compare_other_tensor_libs}}.

\finalresponse{
DISTAL~\cite{DISTAL22} is another recently developed distributed tensor algebra compiler that allows expressing computation on tensors as tensor algebra expressions. DISTAL lets users describe how the data and computation of a tensor algebra expression map onto a target machine allowing separation of the specifications of data distribution and computation distributions. DISTAL lets users specialize computation to the way that data \textblack{are} already laid out or easily transform data between distributed layouts to match the computation. TAMM, on the other hand, offers generality by
decomposing tensor algebra expressions into a series of distributed matrix multiplication and transposition operations. In a more recent work, the authors also introduced SpDISTAL~\cite{spdistal}, which adds support for sparse data structures \textblack{and allows} users to define generic sparse tensors. SpDISTAL then generates code for executing operations over these tensors. Both DISTAL and TAMM have several similarities in the aspects of storage, computation, and scheduling. However, they also differ in certain key aspects, which are further elaborated in Section \ref{sec:feature_compare_other_tensor_libs}.}

\begin{figure}[t]
  \centering
\includegraphics[width=0.5\textwidth]{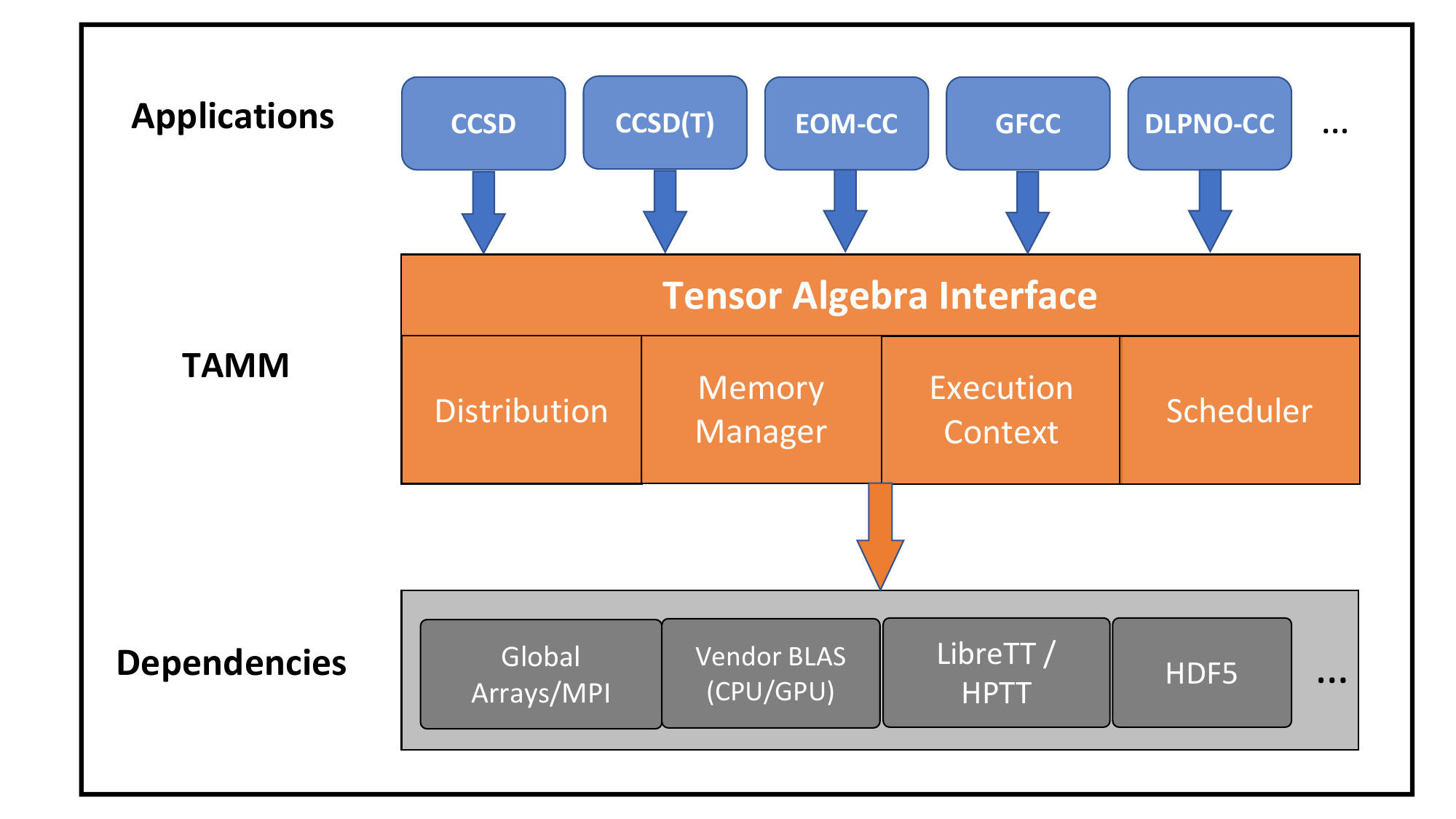} 
  \caption{Overview of tensor algebra for many-body methods (TAMM) framework}
  \label{fig:tamm_overview}
\end{figure}

\section{\textblack{TAMM FRAMEWORK}}
\label{sec:tamm_framework}

This section provides a detailed explanation of our tensor algebra framework. Figure \ref{fig:tamm_overview} shows the conceptual overview of our framework. TAMM provides a tensor algebra interface that allows users to describe tensor operations in a familiar mathematical notation while underneath it employs efficient data distribution/management schemes as well as efficient operation execution on both \textblack{central processing unit (CPU)-}based and GPU-based hardware. Our framework leverages efficient libraries such as HPTT~\citep{hptt_17} for  tensor index permutations on CPUs, LibreTT~\citep{librett} for tensor index permutations on GPUs, and Global Arrays,~\citep{ga_paper} a partitioned global address space (PGAS) programming model for efficient data distribution. 

\subsection{Tensor \textblack{a}lgebra \textblack{o}perations in TAMM} 

TAMM provides a flexible infrastructure for describing tensor objects using the general notion of an \boldit{index space} (simply an index range) that is used to describe the size of each dimension. TAMM also employs tiling capabilities\textblack{,} where users can define arbitrary or fixed size tiling over \boldit{index spaces} to construct \boldit{tiled index spaces} to represent a blocked representation of a tensor. This allows tensors to be constructed as a set of smaller blocks of data indexed by the Cartesian product of the number of tiles on each dimension. This notion of tiling enables efficient data distribution and efficient tensor operations (such as tensor contractions) that can better leverage the underlying hardware by utilizing the available cache system as well as the execution modules (i.e., GPUs \textblack{and} CPUs). TAMM's flexible tiling capabilities provide crucial support for load balancing and making effective trade-offs over data communication and efficient computation. These optimizations will be detailed in \textblack{Subsection \ref{sec:tensor_dist}}, where we describe the data communication and execution of tensor operations.

\begin{figure}[t]
\centering
\begin{lstlisting}
  // Constructing index spaces
  IndexSpace N{range(100)};
  IndexSpace M{range(30)};
  IndexSpace K{range(20)},
               {{"first", {range(0, 10)}},
                {"second", {range(10, 20)}}
               }};

  // Tiling the index spaces
  TiledIndexSpace tN{N, 10};
  TiledIndexSpace tM{M, {10,20}};
  TiledIndexSpace tK{K, 5};

  // Constructing a tensor object
  Tensor <double> A{tM, tK};
  Tensor <double> B{tK, tN};
  Tensor <double> C{tM, tN};

  Tensor<double> ST{{tM, tN}, [] (const IndexVector& block_id, span<T> buf){ 
            /* user-defined logic to fill tensor blocks */ }};
\end{lstlisting}
\caption{Source code example for \code{IndexSpace}, \code{TiledIndexSpace}, and \code{Tensor} construction using TAMM.}
\label{fig:tensor_const}
\end{figure}

Figure \ref{fig:tensor_const} shows an example of \boldit{index space}, \boldit{tiled index space}, and \boldit{tensor} construction in the TAMM framework. Lines 2--4 show the construction of \code{IndexSpace} objects that represent a range of indices using a range constructor. Our framework also allows users to describe application-specific annotations over the subsections of these spaces. Lines 4--7 show an example of string-based annotation over subsections of the construction range. This allows users to easily access different portions of an index space, thereby enabling access to slices of a tensor and constructing new tensors using slices of the original index space. Similarly, users can encode \boldit{spin} information related to each dimension over the input tensors\textblack{,} allowing our run-time to allocate these tensors using a block-sparse representation. \response{For the time being, we only support spin symmetry for representing the block sparsity where the zero-blocks are not stored. Users can define their own non-zero check functionality to be used during allocation and tensor operation execution for storage and computational benefits. Additional capabilities, such as restricted summations or point group and permutational symmetries, are not supported in the tensor specification. Point group symmetry is less concerning since TAMM targets large systems that often do not have such symmetry. However, developers can explicitly code these features using TAMM to construct the desired tensors. This is the case for methods such as CCSD(T), discussed in Section \ref{Sec:Triples}.}
\finalresponse{TAMM also supports specialized tensor construction, called \boldit{lambda tensor}, where the user can provide a C++ lambda expression that specifies how a block of the tensor is computed, as shown in Line 19. This construct can be used to dynamically generate blocks of a tensor as needed. Note that these tensors are not stored in memory; they are read-only objects computed on the fly to be consumed in various tensor operations described later in this section. Similarly, TAMM also provides a special construct called \boldit{view tensor} to describe access to an existing tensor using C++ lambda expressions. The main use of view tensors is to define tensors of different shapes that can \textblack{be used as} a reference tensor for any operation as well as apply possible sparsity mapping/constraints on a dense tensor. Similar to lambda tensors, a view tensor does not allocate any additional storage but rather provides specialized access to the data available in the referenced tensor. We provide details of a use case for view tensors when discussing the DLPNO-CCSD method in Section \ref{sec:dlpno}.}

Lines 10--12 show the construction of \code{TiledIndexSpace} objects that represent a sliced index space that is used in tensor construction to have a blocked structure. Tiling can be applied as a fixed size (i.e., line 10) or as arbitrary tile sizes with full coverage of the index space (i.e., line 11). Finally, lines 15--17 in Figure \ref{fig:tensor_const} show the construction of \code{Tensor} objects using tiled index spaces. Each \code{TiledIndexSpace} object used in the tensor constructor represents a dimension in \textblack{the} constructed tensors. In this example, tensor \code{A} is a $30 \times 20$ matrix with eight blocks\textblack{,} as denoted by \code{tM}\textblack{,} which consists of two tiles of sizes 10 and 20\textblack{,} while \code{tK} consists of four tiles that are size 5. While these lines construct a tensor object, the tensor is collectively allocated by all participating compute nodes in a subsequent operation using a \boldit{scheduler}. 

\begin{figure}[]
  \centering
  \begin{lstlisting}
  // construct tN labels
  auto [a, b, c] = tN.labels<3>();
  // construct tM labels
  auto [i, j, k] = tM.labels<3>();
  // construct tK labels
  auto [l, m, n] = tK.labels<3>();
  // construct subspaces of tK
  auto [x, y, z] = tK("first").labels<3>();
  \end{lstlisting}
  \caption{Source code example for constructing index labels.}
  \label{fig:index_label}
\end{figure}

Another important concept in constructing tensor operations is \boldit{index labels\textblack{,}} \textblack{which allow} specifying tensor operations in familiar mathematical notations and provide slicing capabilities over tensors by using the string-based subsections of the full index space. Labels are associated with \boldit{tiled index spaces} and used in the tensor operation syntax to describe the computation. Depending on the index spaces that the tensors are constructed on, users can specify string-based sub-spaces to define operations on different slices of the underlying allocated tensor object. Figure \ref{fig:index_label} shows examples of \code{TiledIndexLabel} construction using \code{TiledIndexSpace} objects. While lines 1-6 construct the labels over the full spaces, line 8 shows the label creation for the \boldit{first} portion of \textblack{the} \code{tK} index space (see construction on Figure \ref{fig:tensor_const} line 4). These sub-spaces can then be used for specifying operations over sliced portions of the full tensor as long as the index labels are from a subset of the original index space.

\begin{figure}[th]
      \begin{bnf}
        \bnfprod{tensor-op}
          {
            \bnfpn{op-lhs} \bnfts{=} \bnfpn{op-rhs} \\ 
            \bnfalt \bnfpn{op-lhs} \bnfts{+=} \bnfpn{op-rhs}\\
            \bnfalt \bnfpn{op-lhs} \bnfts{-=} \bnfpn{op-rhs}
          }\\
        \bnfprod{op-lhs}
          {\bnfpn{label-tensor}}\\
        \bnfprod{op-rhs}
          {
            \bnfpn{alpha} \\ \bnfalt 
            \bnfpn{label-tensor} \\ \bnfalt
            \bnfpn{alpha} \bnfts{*} \bnfpn{label-tensor} \\ \bnfalt
            \bnfpn{alpha} \bnfts{*} \bnfpn{label-tensor} \bnfts{*} \bnfpn{label-tensor} 
          }\\
        \bnfprod{label-tensor}
          {\bnfpn{tensor-id} \bnfts{(} \bnfpn{label-list} \bnfts{)}}\\
        \bnfprod{alpha}
          {\bnftd{tensor value type}}
      \end{bnf}
  \caption{Tensor operations grammar in extended Backus–Naur form.}
  \label{fig:tensor_ops}
\end{figure}

\finalresponse{TAMM supports several tensor operations: tensor set, tensor addition, subtraction, scale, trace, transpose, general tensor contractions, inner and outer products, and reduction. TAMM supports different tensor data types and several  mathematically valid operations between tensors of different data types (e.g., tensor contraction between a complex and real tensor).}
Figure \ref{fig:tensor_ops} gives the grammar for allowed tensor operations' syntax in the TAMM framework. Each tensor operation syntax rule ($\bnfpn{tensor-op}$) is composed of a left-hand side (lhs, $\bnfpn{op-lhs}$) and a right-hand side (rhs, $\bnfpn{rhs}$) operation. While \boldit{lhs} can only be a labeled tensor construct ($\bnfpn{label-tensor}$), \boldit{rhs} can be of different types that result in different tensor operations\textblack{, as follows}:
\begin{itemize}
  \item alpha value \textblack{[}\code{A(i,l) = alpha}\textblack{]}, which corresponds to a tensor set operation that assigns the corresponding alpha value to all elements of the tensor. 
  \item labeled tensors \textblack{[}\code{A(i,l) +=  alpha * D(l,i)}\textblack{]} correspond to a tensor addition operation (with respect to the label permutation on tensor D) in Eq. (7). 
  \item contraction of two labeled tensors \textblack{[}\code{C(i,a) +=  alpha * A(i,l) * B(l,a)}\textblack{]} updates the lhs with the tensor contraction results in Eq. (8).
\end{itemize}
\finalresponse{Several additional operations on tensors are provided as utility routines in TAMM. Element-wise operations such as square, log, inverse, etc. are provided. Additional useful operations on tensors, such as min, max, norm, etc., are also provided, including high-level routines to read from and write tensors to disk using parallel file I/O. All these operations are defined using a parallel \boldit{block\_for} construct provided by TAMM, allowing users to easily define custom element- and block-wise operations on a tensor. The \boldit{block\_for} construct allows user-defined operations (using C++ lambda expressions) that are executed in parallel on distributed tensors.}

Similar to tensor allocation, all other tensor operations that use and modify the distributed tensor data have to be performed via a \boldit{scheduler} that has \textblack{the} necessary information about the \boldit{execution context} \textblack{of} the operation. \boldit{Execution context} includes required information about the distribution scheme of the tensor data and the execution medium for the operations through \boldit{distribution}, \boldit{memory manager}, and \boldit{process group} constructs. \finalresponse{Note that the tensor operations that are provided as utility routines do not use the scheduler, but still require the \boldit{execution context} to perform the corresponding operation.} TAMM employs a modular construction of these concepts to allow users to implement various distribution schemes as well as different distributed memory frameworks. Figure \ref{fig:tensor_exec} shows the construction of a \code{Scheduler} object using the \boldit{execution context} and \textblack{the} execution of defined tensor operations. After creating a scheduler, users can directly queue tensor operations such as tensor allocation ( line~12), tensor set and add operations (lines~13 \textblack{and} 14), and tensor contractions (line~15). Finally, all queued operations are executed using the \boldit{execution context} on a distributed system. The tensor operations, as well as the operations over the index spaces, are formally described in our previous work.~\cite{mutlu2019toward} The syntax for expressing operations shown in lines~13-15 also indicates the productivity benefits that can be obtained by using TAMM. The operations expressed in these three lines are executed in parallel on CPUs\textblack{,} GPUs\textblack{,} or both on any distributed computing platform. Manually writing parallel code for these operations would lead to a significant increase in the \textblack{number of} source lines of code (SLOC). Extending such manual development of parallel code to a real application with a large number of such operations would only lead to a significant increase (orders of magnitude) in the SLOC count and development time which would also make future improvements to such code infeasible.

\begin{figure}[t]
  \centering
  \begin{lstlisting}
  // Constructing process group, memory manager, distribution 
  // to construct an execution context 
  ProcGroup pg = ProcGroup::create_world_coll();
  ExecutionContext ec{pg, DistributionKind::dense, MemoryManagerKind::ga};
  
  // Constructing a scheduler for executing tensor operations
  Scheduler sch{&ec};
  
  // Tensor operations are queued for execution into the Scheduler
  sch.allocate(A, B, C)
  (A(i, l) = 1.0)                     // SetOp  - Grammar rule (5)
  (B(i, l) = A(i, l))                 // AddOp  - Grammar rule (6)
  (B(i, l) += -1.0 * A(i, l))         // AddOp  - Grammar rule (7)
  (C(i, a) = 0.5 * A(i, l) * B(l, a)) // MultOp - Grammar rule (8)
  // All queued operations are executed.
  .execute();
  \end{lstlisting}
  \caption{Source code example for executing the tensor operations using TAMM.}
  \label{fig:tensor_exec}
\end{figure}

This section \textblack{summarizes} the tensor algebra interface as an embedded domain-specific language (eDSL) in the TAMM framework. By implementing an eDSL, we were able to \textblack{separate} concerns for developing scientific applications. While the high-level tensor abstractions allowed domain scientists to implement their algorithms using a representation close to mathematical formulation, it also allowed framework developers to test various different optimization schemes on the underlying constructs (i.e., different data distribution schemes, operation execution on accelerators, use of different PGAS systems\textblack{,} etc.)\textblack{,} which we will detail in the coming section.

\subsection{Tensor \textblack{d}istribution and \textblack{o}peration \textblack{e}xecution in TAMM}
\label{sec:tensor_dist}

TAMM leverages various state-of-the-art frameworks and libraries to achieve a scalable performance--portable implementation of tensor algebra operations on exascale supercomputing platforms through efficient data distribution and intra-node execution of tensor operation kernels on CPUs and GPUs.
The default \boldit{memory manager} for tensor data distribution in TAMM is based on the Global Arrays framework, a Partitioned Global Address Space (PGAS) programming model that provides a shared memory-like programming model on distributed memory platforms. Global Arrays provides performance, scalability, and user productivity in TAMM by managing the inter-node memory and communication for tensors. A TAMM tensor is essentially a global array with a certain distribution scheme. We have implemented three tensor distribution schemes in TAMM. The first scheme computes an effective processor grid for a given number of processes. A dense tensor is then mapped onto the processor grid. The second scheme is a simple round-robin distribution that allocates equal-sized blocks in a round-robin fashion where the block size is determined by the size of the largest block in the tensor. This distribution over-allocates memory and ignores sparsity. The third scheme allocates the tensor blocks in a round-robin fashion while taking block sparsity into account. By only allocating non-zero blocks in the tensor, it minimizes the memory footprint of overall computation, allowing bigger-sized problems to be mapped to the available resources. 

TAMM uses the ``Single Program Multiple Data (SPMD)'' model for distributed computation. In this programming abstraction, each node has its own portion of tensors available locally as well as access to the remote portions via communication over the network. As a result, operations on whole tensors can result in access \textblack{of} remote portions on the tensor, with implied communication. More importantly, many operations (i.e., tensor contractions, addition, etc.) are implied to be collective as they involve the distributed tensors as a whole. While the tensor algebra interface follows a sequential ordering of tensor operations, we also tried to conceal the burden of thinking in a distributed manner while writing a scientific application. To avoid possible issues with operations on distributed tensors, TAMM is designed to represent tensors in terms of handles and requires tensor operations to be declared explicitly and executed using a \boldit{scheduler}. Hence, any assignment \textblack{performed} on tensor objects will be a shallow copy as opposed to a deep copy, as a deep copy implies communication (message passing between nodes to perform the copy operation). 

The computational chemistry methods are implemented as a sequence of operations over distributed tensors. Through this design, we were able to separate the specification of the operations from their implementations, allowing method developers to mainly focus on the computational chemistry algorithms while kernel developers can focus on the optimization of individual tensor operations. 
\textblack{The execution} of all tensor operations is managed by a scheduler. The TAMM scheduler employs a data flow analysis over the queued tensor operations to find the dependencies between each operation in order to load balance the computation and communication requirements of the overall execution. Using a levelized execution scheme, the scheduler is able to limit the global synchronizations to achieve load-balanced and communication-efficient schedules for the execution of all tensor operations. The dependency analysis over the high-level operations is based on a  \textblack{macro-}operation graph. When two or more operations share the same tensor object and one of these operations updates the shared object, the operations are marked as conflicting operations that can not be executed in parallel. This operation-graph is used to construct a batch of operations that can be executed in parallel, minimizing the total number of global synchronizations required by the computation. The operations in these batches are then executed in an SPMD fashion. For instance, \textblack{the} canonical CCSD implementation in TAMM has 10 levels of operation batches that sum to over 125 tensor operations.

While TAMM hides the burden of choosing the best-performing schedule from the users through load-balanced scheduler execution, it allows users to control various aspects of the computation, such as data distribution, parallelization strategies, operation ordering, and execution medium (i.e., CPUs, GPUs). To achieve this, TAMM uses a modular structure to describe constraints imposed by the users to automate the construction of an execution plan for efficient execution of the computation. This allows users to incrementally optimize their code with minimal programming effort while keeping it readable as opposed to a code generator-based solution.
With these controls over the execution and distribution of the tensor data, users can choose from different optimizations at different \textblack{granularities}. For example, the user can increase the data locality by replicating frequently used small tensors on each distributed node or choosing from different distribution schemes for efficient tensor operation execution on various node topologies.Such an optimization can be implemented as a new data distribution scheme (i.e., SUMMA~\cite{summa}) by extending the \code{Distribution} class. By simply using this distribution, users can enforce a specific distribution on tensors which can optimize required data communication for specific operations. Another abstraction for optimization is the actual computation of the tensor operations. Users can define new operations by extending the \code{Op} class which can then be scheduled to be executed. While TAMM supports the most common operations defined on tensors (i.e., addition \textblack{and} contraction), it also implements a \code{Scan} and \code{Map} operation that can be used to define various element/block-wise operations using lambda functions. Additionally, as a lower-level abstraction, users \textblack{can} also decide to describe new executions specific \textblack{to} a new architecture/accelerator on the kernel level (i.e., DGEMM, GPU abstraction).  While TAMM supports main GPUs (i.e., Nvidia, AMD, \textblack{and} Intel) that will be available in upcoming HPC systems, users can also choose to implement new kernel-level abstractions for different hardware stacks (i.e., FPGAs \textblack{and} ML/DL accelerators).

To achieve highly optimized execution of tensor operations on a given platform, TAMM is designed to allow \textblack{the} use of multiple external libraries that provides optimized kernels for various operations. In addition to leveraging vendor-provided linear algebra libraries that are highly tuned for both CPUs and GPUs, TAMM also uses the following libraries: \textblack{the} HPTT~\cite{hptt_17} library for optimized tensor index permutations on CPUs\textblack{;} LibreTT~\citep{librett} for enabling efficient index permutation on Nvidia, AMD\textblack{,} and Intel GPUs; \textblack{the} BLIS~\cite{blis}library for efficient BLAS implementation in CPU\textblack{;} and \textblack{finally}, TensorGen~\cite{tensorgen} for generating optimized tensor contraction GPU kernels specialized for CC methods in computational chemistry. 

While TAMM tries to hide the execution details from the end-user by employing high-level tensor algebra expressions, users can specify the medium on which they want the operations to be executed. Users can specify a particular operation to be executed on \textblack{the} CPU or GPU. All other execution-specific details like parallelization, \textblack{CPU-to-GPU} communication, and execution mechanisms are handled automatically by \textblack{the} TAMM infrastructure. While TAMM does not explicitly rely on OpenMP, it can leverage OpenMP for important operations like tensor transposition and GEMM via highly optimized external libraries such as HPTT and vendor BLAS. Setting the \code{OMP$\_$NUM$\_$THREADS} variable before running any TAMM-based application would enable OpenMP parallelization.

\subsection{Comparison with other tensor algebra frameworks}
\label{sec:feature_compare_other_tensor_libs}

\begin{center}
\begin{table}[t]
\caption{Feature comparison with other distributed tensor algebra frameworks}
\resizebox{\columnwidth}{!}{%
\begin{tabular}{lcccc}
\hline
\multicolumn{1}{c}{\textbf{Frameworks}} & \textbf{Indexing}                  & \textbf{Slicing}         & \textbf{GPU \textblack{s}upport} & \textbf{Programming \textblack{m}odel} \\ \hline
\textbf{TAMM}                           & Object- \textblack{and} string-based             & Generic slicing          & AMD, Intel, Nvidia   & GA (\textblack{e}xtendable)        \\
\textbf{TA}                             & String-based                       & No support               & Nvidia               & MADNESS          \\
\textbf{DISTAL}                         & {\color{black}Object-based}                       & No support               & Nvidia               & Legion           \\
\textbf{CTF}                            & String-based                       & No support               & Nvidia               & MPI              \\
\textbf{ExaTensor}                      & Object-based                       & Sub-range slicing        & AMD, Nvidia          & MPI              \\ \hline
\end{tabular}%
}
\label{tab:feature_comp}
\end{table}
\end{center}

\finalresponse{In this section, we compare the functionality provided by TAMM with \textblack{that provided by} other distributed tensor algebra frameworks - TiledArray, CTF, DISTAL, and ExaTensor. Table~\ref{tab:feature_comp} provides the key features \textblack{and} differences between these frameworks. While all these frameworks provide similar features in how tensors and tensor operations can be represented, TAMM is designed to be an extendable framework and hence differs from the remaining frameworks in the following key aspects discussed below\textblack{:}}

\begin{itemize}
\item \finalresponse{\textbf{Indexing and slicing:} Indexing and slicing capabilities in any tensor algebra framework are crucial to \textblack{defining} tensor operations. Indexing plays a key role in constructing tensor operations (see Figure \ref{fig:tensor_exec}). Slicing is predominantly used in computational chemistry methods\textblack{,} allowing operations on isolated parts of tensors. Users can choose to write the tensor operations over slices of the tensors instead of having to copy \textblack{the} required portions first for such operations. While the tensor algebra frameworks being compared provide an indexing mechanism, they lack support for using slicing in the tensor operations. TA and CTF use a string-based solution for indexing but lack the capability to support sliced indexing. On the other hand, {\color{black}DISTAL and} ExaTensor use object-based constructs to create and use labels in the computation. {\color{black} While DISTAL does not allow any slicing, ExaTensor let users to use pre-defined sub-ranges for slicing}. TAMM provides generic indexing and slicing capabilities through various index space operations allowing users to define a subset of indices to create object-based labels (see Figure \ref{fig:index_label}) as well as string-based labels to construct tensor operations. Users can make use of slicing capabilities by using object-based labels {\color{black} over sub-ranges or by creating new slices on-the-fly} or use string-based labels to define operations over full tensors.}

\item \finalresponse{\textbf{Sparsity:} Sparsity is widely used in several computational chemistry applications. All of the tensor algebra frameworks discussed in this section support some notion of sparsity for both data and computation. While DISTAL does not have direct support for sparsity, later work by the same authors, namely SpDISTAL, incorporates different sparse data structures as well as specifications to describe the storage and computation \textblack{of} sparse tensors. SpDISTAL supports a generic specification for describing {\color{black} the dense/compressed format for each individual mode} as well as pre-defined sparse data representations (i.e., CSR, CSC etc.) for describing such tensors. Operations over these tensors are executed by \textblack{generating} custom kernels. To contrast, more application-oriented tensor libraries, such as TA, CTF, and ExaTensor, allow block-level sparsity. TA allows users to define non-zero and zero tiles that construct the tensors, whereas CTF uses pre-defined symmetry-based sparsity over the tensors. TAMM\textblack{,} on the other hand\textblack{,} incorporates sparsity by providing attributes over the tiled index spaces. By default, TAMM implements spin symmetry-related attributes to allocate block sparse tensors depending on the values of corresponding attributes on the tiles of the index space. Internally, TAMM analyzes the tensor blocks to determine if the corresponding block is non-zero to optimize storage and computation on block sparse tensors.}

\item \finalresponse{\textbf{GPU Support:} It is crucial to effectively leverage the available resources in modern  heterogeneous HPC platforms. TAMM currently enables tensor algebra operations (mainly tensor contractions) on multiple GPU architectures (AMD, Nvidia, \textblack{and} Intel). At the time of this writing, the remaining frameworks are limited to Nvidia GPUs (except ExaTensor\textblack{,} which has AMD GPU support). The unified GPU infrastructure design allows TAMM to be easily extended to add support for a new hardware accelerator.
Support for the upcoming GPU architectures with performance portability was also explored using SYCL, a domain-specific, heterogeneous programming language. Using SYCL and oneAPI Math Kernel Library (oneMKL) interfaces,\cite{onemkl} perturbative triples contribution from \textblack{the} CCSD(T) formalism was demonstrated with large-scale simulations on different GPU architectures.\cite{sycl_portability}}

\item \finalresponse{\textbf{Extendable backends:} TAMM currently uses Global Arrays (GAs)\textblack{,} which is built on top of MPI\textblack{,} for managing all aspects of data distribution and communication. TAMM also has an experimental UPC++~\cite{upc++} backend that can alternatively be used instead of GA. 
TiledArray(TA) uses \textblack{the} MADNESS~\cite{madness_16} parallel runtime as the backend\textblack{;} DISTAL uses \textblack{the} Legion~\cite{legion} programming model and runtime system for distributing the data and \textblack{computing;} CTF and ExaTensor use MPI as the underlying parallel programming model. TAMM is designed to be easily extendable by implementing the methods in the \boldit{process group} and \boldit{memory manager} classes to support other parallel programming models. To the best of our knowledge, the remaining frameworks are not easily extendable to add support for additional parallel programming models.} 

\item 
\finalresponse{\textbf{Distribution:} A TAMM tensor is allocated with a distribution type specified  via the \boldit{execution context}. The distribution choices that are supported can be extended via the \boldit{distribution} class. DISTAL allows \textblack{the} specification of tensor distributions using a notation that allows the user to specify a process grid that the dimensions of a tensor can be mapped onto. Tensor distributions in TA and CTF are based on several parallel matrix multiplication \textblack{algorithms} (SUMMA, 2.5D, \textblack{and} 3D) ~\cite{summa_16}. {\color{black} TA additionally gives users the option of specifying a process map onto which tiles are distributed. However, CTF} cannot be easily extended to allow other distributions.}

\item
\finalresponse{\textbf{Scheduling and Execution:} DISTAL allows users to specialize computation to the way that data \textblack{are} already laid out or easily transform data between distributed layouts to match the computation\textblack{,} allowing fully customizable execution strategies for each operation. Several parallel matrix multiplication algorithms from \textblack{the} literature~\cite{summa_16} are expressible as schedules in DISTAL, which is currently not possible with any other framework in comparison. TA and CTF automatically execute tensor operations by dynamically redistributing data across operations if needed but do not provide any way to \textblack{customize} the operation scheduling process. TAMM allows deferring execution for a collection of tensor operations.  When the user explicitly calls the \boldit{execute} function, the TAMM scheduler analyzes \textblack{the} dependencies between the various operations, and schedules them appropriately in order to load balance the computation and communication requirements of the overall execution across the collection of operations. However, TAMM does not provide any further \textblack{customization} to the scheduling process. The trade\textblack{-}off between these frameworks is that TAMM, TA, and CTF fully automate the distribution process, while users must explicitly provide a schedule to distribute their computations with DISTAL.}

\end{itemize}

In this section, we have provided a detailed explanation of the TAMM framework. TAMM leverages a modular infrastructure to enable the implementation of various optimizations on different levels of computation, from data distribution to execution schemes on different hardware. This design allowed us to implement different \boldit{memory managers}, \boldit{distribution} schemes, and work distribution over different \boldit{process groups} without any major changes to the user-facing tensor algebra interface. \finalresponse{Finally, we provided a feature-based comparison between TAMM and other
distributed tensor algebra frameworks, namely TiledArray, CTF, DISTAL, ExaTensor.}

\section{\textblack{CC TAMM IMPLEMENTATIONS}}
\label{sec:case_studies}

In this section, we present case studies where TAMM was used to implement various scalable coupled-cluster (CC) methods for the latest HPC systems. While important, TAMM's primary contributions are not just the faster performing versions of these methods but \textblack{also} the ability to productively develop and explore new algorithms, and apply those improvements to all existing and new applications implemented using TAMM. This includes improvements in intra-node execution (single core, OpenMP multicore, GPUs, etc.), data distribution strategies (e.g., replication, process group-based distribution), and parallel execution (compute partitioning and communication scheduling algorithms). It also allows the concurrent development of optimized equations, parallel algorithms, and optimized intra-node kernels by different teams through clearly defined interfaces. The variety of methods presented and their performance \textblack{are} evidence that this approach accelerates the development and exploration of CC methods.

\subsection{Canonical Methods}

The canonical formulations of the Coupled-Cluster (CC) formalisms
\cite{coester58_421,coester60_477,cizek66_4256,paldus72_50,purvis82_1910,paldus07,bartlett_rmp}
are based on the exponential parametrization of the correlated  ground-state wave function $|\Psi\rangle$:
\begin{equation}
|\Psi\rangle = e^T |\Phi\rangle \;,
\label{cc1}
\end{equation}
where $T$ is the cluster operator and the reference function $|\Phi\rangle$ in single-reference formulations is assumed to be represented by the Hartree$-$Fock Slater determinant. In practical realizations, it is assumed that $|\Phi\rangle$ provides a good approximation to the correlated ground state, $|\Psi\rangle$.
The cluster operator $T$ can be partitioned into it many body components $T_k$
\begin{equation}
T=\sum_{k=1}^{N} T_k \;,
\label{cc2}
\end{equation}
defined as 
\begin{equation}
T_k = \frac{1}{(k!)^2} \sum_{i_1,\ldots,i_k; a_1,\ldots, a_k} t^{i_1\ldots i_k}_{a_1\ldots a_k} 
a^{\dagger}_{a_1} \ldots a^{\dagger}_{a_k} a_{i_k} \ldots a_{i_1}\;,
\label{cc3}
\end{equation}
where $a_p^{\dagger}$ ($a_p$) are the creation (annihilation) operators for an electron in $p$-th state and indices $i_1,i_2,\ldots$ ($a_1,a_2,\ldots$) refer to occupied (unoccupied) spin orbitals in the reference function $|\Phi\rangle$. The operators $T_k$, defined by the 
cluster amplitudes $t^{i_1\ldots i_k}_{a_1\ldots a_k}$, produce $k$-tuple excitations  when acting onto the reference function.

To define \textblack{the }equations needed for determining cluster amplitudes, we introduce wave-function expansion (\ref{cc1}) 
into the Schr\"odinger equation \textblack{by} pre-multiplying both sides by $e^{-T}$. 
This procedure leads to an explicitly connected form of the energy-independent 
equations for amplitudes and energy, i.e., 
\begin{eqnarray}
\langle\Phi_{i_1\ldots i_k}^{a_1 \ldots a_k}|e^{-T}He^{T}|\Phi\rangle = 0\;, &&\forall_k, \forall{i_1,\ldots,i_k},
\forall{a_1,\ldots,a_k}\;, \label{cc4} \\
E=\langle\Phi|e^{-T}He^{T}|\Phi\rangle \;, && \label{cc5}
\end{eqnarray}
where the electronic Hamiltonian $H$ is defined as
\begin{equation}
    H=\sum_{pq} h^{p}_{q} a_q^{\dagger} a_p + \frac{1}{4} \sum_{p,q,r,s} v^{pq}_{rs} a_r^{\dagger} a_s^{\dagger} a_q a_p 
    \label{cc6}
\end{equation}
where $h^p_q$ and $v^{pq}_{rs}$ are tensors representing interactions in the quantum system and excited Slater determinants 
$|\Phi_{i_1\ldots i_k}^{a_1 \ldots a_k}\rangle$ are defined as
\begin{equation}
|\Phi_{i_1\ldots i_k}^{a_1 \ldots a_k}\rangle = 
E^{a_1\ldots a_k}_{i_1\ldots i_k} |\Phi\rangle \;.
\label{exslp}
\end{equation}
In all CC formulations discussed here, to form  equations (\ref{cc4}) for cluster amplitudes, one needs to (1) find and efficient way for  distributing and compressing tensors  $h^p_q$, $v^{pq}_{rs}$, \textblack{and} $t^{i_1\ldots i_k}_{a_1\ldots a_k} $ across all nodes, (2) define efficient algorithms for partitioning TCs of multi-dimensional tensors across the parallel system, and
(3) optimize communication between nodes to minimize the effect of the latency.
To illustrate the scale of the problems, in brute force simulations, we have to store \textblack{four}-dimensional tensors $t^{i_1 i_2}_{a_1 a_2}$ and  $v^{pq}_{rs}$ that require 
storage proportional to $n_o^2 n_u^2$ and $(n_o+n_u)^4$, respectively, where 
$n_o$ and $n_u$ refer to the number of occupied and unoccupied orbitals in the reference function $|\Phi\rangle$. For CC simulations of the molecular systems defined by $n_o=200$ and $n_u=2,800$ one needs to distribute data of the order of 150 TB.
Therefore, to make these simulations possible, the support \textblack{of} sophisticated HPC algorithms and applied math is indispensable.

\subsubsection{Coupled \textblack{c}luster \textblack{s}ingles \textblack{d}oubles (CCSD)}\label{ccsd_section}

The CCSD formalism (CC with single and double excitations) 
\cite{purvis82_1910}
is one of the most popular CC approximations and is used in routine computational chemistry calculations as a necessary intermediate step toward more accurate CC models, such as \textblack{the} perturbative CCSD(T) formalism discussed in \textblack{Section \ref{Sec:Triples}}, or excited-state or linear-response CC extensions. In the CCSD formalism, the cluster operator $T$ is approximated as
\begin{equation}
T \simeq T_1 + T_2 \;,
\label{cc7}
\end{equation}
and the equations for cluster amplitudes are represented as 
\begin{eqnarray}
r_i^a &=& \langle\Phi_i^a| e^{-(T_1+T_2)} H e^{T_1+T_2}|\Phi\rangle \;, \label{cc8a} \\
r_{ij}^{ab} &=& \langle\Phi_{ij}^{ab}| e^{-(T_1+T_2)} H e^{T_1+T_2}|\Phi\rangle \;, \label{cc8b}
\end{eqnarray}
where the tensors $r_i^a$ and $r_{ij}^{ab}$ are commonly referred to as the residual vectors. 
Due to \textblack{the} large number of terms corresponding to complicated contractions between $h^p_q$/$v^{pq}_{rs}$ and $t^i_a$/$t^{ij}_{ab}$, optimization of the expressions plays a crucial role. This is achieved by proper factorization by introducing the so-called  recursive intermediates. 
For example, 
\begin{equation}
  \frac{1}{4} v^{ef}_{mn} t^{ij}_{ef} t^{mn}_{ab}
  \label{cc9}
\end{equation}
term, which contributes to $r_{ij}^{ab}$ in the naive approach is characterized by $n_o^4 n_u^4$ numerical overhead. However, 
by introducing the intermediate tensor $I^{ij}_{mn}$
defined as (we assumed Einstein summation convention over repeated indices)
\begin{equation}
    I^{ij}_{mn}= v^{ef}_{mn} t^{ij}_{ef}\;,
    \label{cc10}
\end{equation}
the term (\ref{cc9}) can be given by the equation
\begin{equation}
    \frac{1}{4} I^{ij}_{mn} t^{mn}_{ab}
    \label{cc11}
\end{equation}
at the total numerical cost proportional to $n_o^4 n_u^2$. 

\finalresponse{
Equation ~\ref{cc11} can be expressed using the TAMM interface as follows,
\begin{eqnarray}
(I(m,\, n,\, i,\, j\,)   &=& v(m,\, n,\, e,\, f\,)*t(e,\, f,\, i,\, j\,)) \nonumber \\
(r(a,\, b,\, i,\, j\,) &+=& 0.25*I(m,\, n,\, i,\, j\,)*t(a,\, b,\, m,\, n\,))
\label{tamm_cc11}
\end{eqnarray}
The TAMM code in (\ref{tamm_cc11}) encapsulates distributed execution of the tensor contractions on CPUs, GPUs as well as the choice of the parallel programming model used by TAMM for managing distributed tensor data. Similarly, all other CCSD equations can be represented as tensor operations using the TAMM interface. As indicated in (\ref{tamm_cc11}), the total number of lines of code is greatly reduced in comparison to hand-coded implementations where one needs to explicitly manage data distribution using a particular programming model and also aspects of execution on various heterogeneous architectures. On the other hand, the TAMM representation of such equations makes it easy to develop and maintain scalable electronic structure methods (such as CCSD) that can be run on a variety of HPC platforms.
}

Another important technique that is useful in reducing the memory footprint of the CCSD approach is the Cholesky decomposition (CD) of the $v^{pq}_{rs}$ tensor, i.e.,
\begin{equation}
    v^{pq}_{rs}\simeq (pr|L)(L|qs)-(ps|L)(L|qr) \;,
    \label{cc12}
\end{equation}
where $L$ is an auxiliary index.
The total memory required to store Cholesky vectors $(pq|L)$ needed to reproduce $v^{pq}_{rs}$
with high accuracy is usually proportional to $(n_o+n_u)^3$. 

We developed both closed-shell and open-shell implementations of the CCSD equations using TAMM. The use of TAMM \boldit{index spaces} for alpha and \textblack{beta-}occupied and virtual orbitals allowed us to represent the same theory optimized for specific problems easily. An example equation that is a contribution to the intermediate $I$ in Eqn.~\ref{tamm_cc11} may \textblack{appear} like the following:
\begin{eqnarray}
&&(I(m\_a,\, n\_a,\, i\_a,\, j\_a\,)   =\nonumber \\
&&v(m\_a,\, n\_a,\, e\_b,\, f\_b\,)*t(e\_b,\, f\_b,\, i\_a,\, j\_a\,))\,\,\, ,
\label{tamm_cc-Spin}
\end{eqnarray}
where the added $\_a$ and $\_b$ notations respectfully refer to the corresponding indices for the alpha and beta sub-spaces. While the equations for the closed- vs open- shell implementations are different, the ease of writing them is the same due to the slicing capabilities on sub-spaces provided by TAMM.


As  benchmark systems for testing the performance of TAMM implementations of the Cholesky-based CCSD\cite{ccsd_github},  we used two  molecular systems previously used in studying the efficiency of the TCE implementations of the CC methods in NWChem \cite{sc2011kk, hu2014toward}. The first benchmark system  is the  model of Bacteriochlorophyll (BChl)  MgC$_{36}$N$_4$O$_6$H$_{38}$, \cite{bchl1,bchl2} which plays an important role in understanding the mechanism of photosynthesis. In this process, the light is harvested by antenna systems and further funneled to BChl molecules, \textblack{which initiate} primary charge separation.  The second benchmark system considered here is the $\beta$-carotene molecule, whose doubly excited states and triplet electronic states have recently been intensively studied in the context of singlet fission processes in carotenoid aggregates.\cite{beta1,beta2,beta3}
The $\beta$-carotene molecule consists of 96 atoms, while the BChl model contains 85 atoms, including a central magnesium atom. We use the cc-pVDZ basis for both systems and evaluate the\textblack{ir} performance on OLCF Summit.~\cite{olcfsummit} Time per CCSD iteration is given. Each Summit node contains two IBM POWER9 processors, each consisting of 22 cores and 512 GB of CPU memory. Each node is also equipped with \textblack{six} Nvidia Volta GPUs, each with 16 GB \textblack{of} memory, \textblack{for} a total of 96 GB \textblack{of} GPU memory per node.

Table~\ref{tab:ccsd_performance} shows the CCSD performance compared to NWChem on 200 nodes of OLCF Summit. We measure the performance of the TAMM implementations against the TCE implementations in NWChem.
\textblack{The time} per CCSD iteration is given in seconds. NWChem has \textblack{a} CPU-only implementation and uses 42 CPU cores on each node of Summit. TAMM-based Cholesky-CCSD uses only \textblack{six} MPI ranks per node, where each MPI rank is mapped to a single GPU. For these two molecular systems, we observe a 9-15$\times$ speedup with the TAMM-based Cholesky-CCSD implementation compared to the TCE CCSD method in NWChem. The CPU implementation of the CCSD tensor operations in NWChem comprises 11,314 source lines of code whereas the Cholesky-CCSD implementation expressed using the TAMM framework is only 236 source lines of code. Since these 236 lines represent computation at a high level, they express both CPU and GPU computation. On the other hand, adding GPU capabilities to the NWChem CCSD code will only significantly increase the SLOC count and development time which is why a GPU implementation for CCSD \textblack{has not been} attempted to date in NWChem. This clearly demonstrates the productivity benefits of expressing such computations in TAMM. \response{Our implementation of the CCSD equations~\cite{ccsd_github} \textblack{is} expressed similar\textblack{ly} to as the example in Figure \ref{fig:tensor_exec}. CCSD is an example of how TAMM can be used to productively create an effective and efficient implementation of certain classes of computational chemistry methods. The remaining computational chemistry methods discussed in this paper are also implemented along similar lines.}
 
\begin{center}
\begin{table}
\centering
\caption{TAMM performance compared to NWChem on 200 nodes of OLCF Summit. Time per CCSD iteration is given.}
\begin{tabular}{lccccc}
 \hline
 \hline
 Molecule & \#Atoms &\#Occupied& \#Basis & \multicolumn{2}{c}{Time (s)} \\
 \cline{5-6}
        &         & Orbitals & Functions       & NWChem\cite{apra2020nwchem}           & TAMM      \\ \hline
 \hline
 BChl & 85 & 171 & 852 & 1202 &  81 \\ 
 $\beta$-carotene & 96 & 148 & 840 & 801  &  65  \\
\hline
\end{tabular}
\label{tab:ccsd_performance}
\end{table}
\end{center}


\subsubsection{Coupled \textblack{c}luster \textblack{t}riples}\label{Sec:Triples}

The CCSD(T) formalism 
\cite{ccsd_t,stanton1_t}
is capable in many cases of providing the so-called chemical level of accuracy required in studies of chemical reactivity and thermochemistry. 
In the CCSD(T) approach the perturbative correction 
due to triple excitations ($E^{(\rm T)}$) is added to the CCSD energy:
\begin{equation}
E^{\rm CCSD(\rm T)} = E^{\rm CCSD} + 
E^{(\rm T)} \;,
\label{cc13}
\end{equation}
where 
\begin{eqnarray}
E^{(\rm T)}  &=& \sum_{\substack{i<j<k \\ a<b<c}}
\frac{\langle\Phi|(T_2^+V_N)|\Phi_{ijk}^{abc}\rangle\langle\Phi_{ijk}^{abc}|V_NT_2|\Phi\rangle}
{\epsilon_i+\epsilon_j+\epsilon_k-\epsilon_a-\epsilon_b-\epsilon_c}
\nonumber \\
&& \hspace*{-0.4cm} +\sum_{\substack{i<j<k \\ a<b<c}}
\frac{\langle\Phi|T_1^+V_N|\Phi_{ijk}^{abc}\rangle\langle\Phi_{ijk}^{abc}|V_NT_2|\Phi\rangle}
{\epsilon_i+\epsilon_j+\epsilon_k-\epsilon_a-\epsilon_b-\epsilon_c} \;,
\label{cc14}
\end{eqnarray}
where $V_N$ is the two-body part of the electronic Hamiltonian in a normal  product form, and
$|\Phi_{ijk}^{abc}\rangle= a_a^{\dagger} a_b^{\dagger} a_c^{\dagger} a_k a_j a_i|\Phi\rangle $.
The most expensive part of the CCSD(T) calculation,
characterized by the $n_o^4 n_u^3 + n_o^3 n_u^4$ scaling, 
is associated with calculating the
$\langle\Phi_{ijk}^{abc}|V_NT_2|\Phi\rangle$ term,
which is defined as 
\begin{eqnarray}
&&\langle\Phi_{ijk}^{abc}|V_NT_2|\Phi\rangle = \nonumber \\
   && v^{ij}_{ma} t^{mk}_{bc} 
    - v^{ij}_{mb} t^{mk}_{ac}
    + v^{ij}_{mc} t^{mk}_{ab}
    - v^{ik}_{ma} t^{mj}_{bc} 
    + v^{ik}_{mb} t^{mj}_{ac} + \nonumber \\
&&    v^{ik}_{mc} t^{mj}_{ab} 
    + v^{jk}_{ma} t^{mi}_{bc} 
    - v^{jk}_{mb} t^{mi}_{ac} 
    + v^{jk}_{mc} t^{mi}_{ab} 
    - v^{ei}_{ab} t^{jk}_{ec} + \nonumber \\
&&    v^{ei}_{ac} t^{jk}_{eb}
    - v^{ei}_{bc} t^{jk}_{ea} 
    + v^{ej}_{ab} t^{ik}_{ec} 
    - v^{ej}_{ac} t^{ik}_{eb}
    + v^{ej}_{bc} t^{ik}_{ea} - \nonumber \\
&&  v^{ek}_{ab} t^{ij}_{ec} + v^{ek}_{ac} t^{ij}_{eb}
-v^{ek}_{bc} t^{ij}_{ea},
~~(i<j<k, a<b<c).
\label{tensort}
\end{eqnarray}
Eq. (\ref{tensort}) 
can be separated into terms $A^{abc}_{ijk}$ and $B^{abc}_{ijk}$,
defined by contractions over occupied indices  
($A^{abc}_{ijk}$; \textblack{the} first  nine  terms  on  the  right  hand  side  (r.h.s.) of  Eq. (\ref{tensort}))  and  terms  corresponding  to  contraction  over unoccupied indices  \textblack{[}$B^{abc}_{ijk}$;  remaining  nine  terms  on  the  r.h.s. of Eq. (\ref{tensort})\textblack{]}, i.e.,
\begin{equation}
\langle\Phi_{ijk}^{abc}|V_NT_2|\Phi\rangle =
A^{abc}_{ijk}+B^{abc}_{ijk} \;.
\label{abt}
\end{equation}
Analogously, the $\langle\Phi_{ijk}^{abc}|V_NT_1|\Phi\rangle$ term takes the form 
\begin{eqnarray}
\langle\Phi_{ijk}^{abc}|V_NT_1|\Phi\rangle &=& +v_{ab}^{ij} t^k_c - v_{ac}^{ij}t^k_b +v_{bc}^{ij}t_a^k \nonumber \\
&& -v_{ab}^{ik} t_c^j+v_{ac}^{ik} t_b^j -
v_{bc}^{ik} t_a^j \nonumber \\
&& +v_{ab}^{jk} t_c^i - v_{ac}^{jk} t_b^i 
+ v_{bc}^{jk} t_a^i \nonumber \\
&& (i<j<k, a<b<c)  \;.
\label{tensort2}
\end{eqnarray}
\response{While Eqs. (\ref{tensort}) and (\ref{tensort2}) can be easily implemented with the tensor operations mentioned  in {\color{black}Sec. II A}, the performance of the (T) correction, in terms of both speed and memory,  is improved by encoding key symmetries by hand while still using the distributed tensor data structure provided by TAMM. This includes restricted index summation and permutational symmetries, which are not routine features in the current version of TAMM. In addition,  the contractions in (T) are all fused so that the resulting output is a scalar instead of a \textblack{six}-dimensional intermediate tensor. Currently, TAMM does not support fusion across contractions writing to the same output tensor, but this capability is considered for the near future. Furthermore, the hand-coded kernels for (T) for various GPU architectures are required for the best performance. In addition to fusion, we are also considering incorporating optimized GPU kernels in TAMM for specific contractions such as the ones in (T). The scheduler can then detect such contractions and choose the corresponding optimized kernels for execution rather than executing the contractions through the current default pipeline.}

Table~\ref{tab:ccsd_t_performance} shows the TAMM-based {\color{black}NWChemEx} triples correction (T) calculation\cite{ccsd_github} performance compared to NWChem on 512 nodes of OLCF Summit. Time is given in seconds. NWChem has a GPU implementation of the triples correction and uses all 6 GPUs and 42 CPU cores on each node. The TAMM-based triples correction uses only \textblack{six} MPI ranks per node, where each MPI rank is mapped to a single GPU. For the BChl and $\beta$-carotene molecules, we observe a speedup of $\sim$6-18$\times$ for the TAMM implementation compared to the TCE implementation in NWChem. The finer details of the TAMM-based triples correction implementation are detailed in Ref.~\citenum{triples_sc20}.
\begin{center}
\begin{table}[h]
\centering
\caption{{\color{black}NWChemEx-}TAMM performance compared to NWChem for the perturbative correction in the CCSD(T) method on 512 nodes of OLCF Summit.}
\begin{tabular}{lccccc}
 \hline
 \hline
 Molecule & \#Atoms &\#Occupied& \#Basis & \multicolumn{2}{c}{Time (s)} \\
 \cline{5-6}
        &         & Orbitals & Functions       & NWChem\cite{apra2020nwchem}       & TAMM      \\ \hline
 BChl & 85 & 171 & 852 & 18285 &  1791 \\ 
 $\beta$-carotene & 96 & 148 & 840 & 6646  &  1164  \\
 \hline
\end{tabular}
\label{tab:ccsd_t_performance}
\end{table}
\end{center}


\subsection{New \textblack{m}ethods}

\subsubsection{Equation-of-\textblack{m}otion \textblack{c}oupled \textblack{c}luster \textblack{f}ormalism}
The  equation-of-motion (EOM) methods 
\cite{bartlett89_57,bartlett93_414,stanton93_7029,xpiecuch,kkppeom}
and closely related linear response (LR)  CC formulations 
\cite{monkhorst77_421,jorgensen90_3333,koch97_1808}
can be viewed as excited-state extensions of the single-reference CC theory. In the exact EOMCC formalism, the wave function for the K-th excited state $|\Psi_{\rm K}\rangle$  is represented as
\begin{equation}
|\Psi_{\rm K}\rangle = R_{\rm K} e^T |\Phi\rangle \;,
\label{eomcc1}
\end{equation}
where $R_{\rm K}$ is the excitation operator, which produces the K-th excited state when acting onto the correlated ground-state in CC representations. The energy of the K-th state and the amplitudes defining $R_{\rm K}$ operator can be calculated by solving \textblack{non-Hermitian} eigenvalue problem 
\begin{equation}
\bar{H} R_{\rm K}|\Phi\rangle = E_{\rm K} R_{\rm K}|\Phi\rangle \;,
\label{eomcc2}
\end{equation}
where the similarity transformed Hamiltonian $\bar{H}$ is defined as
\begin{equation}
\bar{H}=e^{-T}H e^T \;, 
\label{eomcc3}
\end{equation}
In the rudimentary EOMCCSD approximation (EOMCC with singles and doubles), the 
$R_{\rm K}$ and $T$ operators are approximated as 
\begin{eqnarray}
R_{\rm K} &\simeq& R_{{\rm K},0}+R_{{\rm K},1}+ R_{{\rm K},2} \;, \label{eomcc4} \\
T &\simeq& T_1+T_2 \;. 
\label{eomcc5}
\end{eqnarray}




\finalresponse{Similar to the CCSD method, the numerical scaling of the EOM-CCSD approach is dominated by $n_o^2 n_u^4$ terms, where $n_o$ represents the number of occupied orbitals and $n_u$ represents the number of unoccupied orbitals. The major computational complexity in the EOMCC method arises from tensor contractions involving various operators. The method involves two types of excitation operators: the CC operator $T$ and the EOM operator $R$, besides the one-body and two-body components of the Hamiltonian operator. The EOMCCSD approach is a non-Hermitian eigenvalue problem, and because of the dimensionality of the problems, \textblack{they are} solved iteratively using a multi-root solver. \textblack{Therefore,} there is an additional cost for constructing and working with the iterative subspace. In terms of implementation, however, the same TAMM format used for writing various tensor contraction equations, as demonstrated while discussing Equation \ref{cc11}, was followed for the EOMCC method.}



The EOMCCSD method is usually employed in studies of excited states dominated by single excitations. It is also worth mentioning that LR-CC methods with singles and doubles leads to the same values of excited-state energies. However, in contrast to the EOMCCSD formalism, the LR-CCSD excitations are identified with the poles of frequency-dependent linear-response CC amplitudes. 

Table~\ref{tab:eomccsd_performance} shows the TAMM EOMCCSD calculation performance compared to NWChem on 200 nodes of OLCF Summit. \textblack{The time} per EOMCCSD iteration is given in seconds. NWChem has CPU-only implementation and uses 42 CPU cores on each of the 200 nodes of Summit. Just as with the previous calculations, the TAMM-based EOMCCSD uses 6 MPI ranks per node, where each MPI rank is mapped to a single GPU. \response{The current TAMM implementation of the EOMCCSD approach has not been optimized at the \textblack{same} equation level as the CD-CCSD implementation. For example, in contrast to the CD-CCSD formalism, the EOMCCSD implementation fully represents \textblack{two}-electron integrals and uses a non-spin-explicit representation of the operators in the equations. While we plan to return to this implementation and incorporate the optimized (spin-explicit) equations, we already observe a significant speed-up over NWChem with this primitive implementation. CD-CCSD and EOMCCSD calculations are different algorithms for solving the corresponding problem, and the timing for an iteration of EOMCCSD is a sum of all of the steps in an interaction, which tends to be more computationally demanding than the Jacobi step in CD-CCSD.} Nonetheless, one can see from Table \ref{tab:eomccsd_performance} that the TAMM EOMCCSD code is 2-3 times faster for $\beta$-carotene and BChl molecules. 

\begin{center}
\begin{table}
\centering
\caption{TAMM performance compared to NWChem on 200 nodes of OLCF Summit. Time per EOMCCSD iteration is given.}
\begin{tabular}{lccccc}
 \hline
 \hline
 Molecule & \#Atoms &\#Occupied& \#Basis & \multicolumn{2}{c}{Time (s)} \\
 \cline{5-6}
        &         & Orbitals & Functions &  NWChem\cite{apra2020nwchem}    &  TAMM         \\  \hline
 BChl & 85 & 171 & 852 & 2030 & 715  \\ 
 $\beta$-carotene & 96 & 148 & 840 & 1170 &  540  \\
 \hline
\end{tabular}
\label{tab:eomccsd_performance}
\end{table}
\end{center}

\subsubsection{DLPNO CCSD(T)}
\label{sec:dlpno}

The development of reduced-scaling quantum chemical methods became a significant part of the recent research effort. In particular, for coupled cluster approaches, the local domain-based methods have been introduced and successfully applied on large systems.
Using even small computational clusters, the 
(DLPNO-CC)\cite{riplinger2013efficient,dlpno-sparse,dlpno-sparse2,pavosevic2016,saitow2017new,10.1039/d1sc03868k} formalism can be \textblack{used} for systems described by 10,000-40,000 basis set functions. The limit for system size can be further extended by utilizing parallel exascale architectures. However, to achieve this goal, several computational challenges associated with the data representation, distribution, and optimization of the DLPNO-CC equations (characterized by a large number of contractions and tensors involved in the underlying equations)  have to be appropriately addressed. Our implementation is inspired by the work cited \textblack{in} Ref.~\citenum{dlpno-sparse2}.
The basic idea is to carefully take advantage of the local character of \textblack{correlational} effects. The total CCSD energy can be seen as a sum of $ij$-pair specific contributions $\epsilon_{ij}$: 
\begin{equation}
E^{\rm CCSD} = \sum_{ij} \epsilon_{ij} \;.
\label{ccsd_en}
\end{equation}
In order to significantly reduce scaling, we are considering \textblack{the} following tasks {\color{black}for the NWChemEx DLPNO CCSD(T) implementation using TAMM}:

\begin{enumerate}
\item Differentiate pairs with respect to their energy contributions $\epsilon_{ij}$ by sequential pre-screenings. Identify $ij$-pairs \textblack{that} a) are negligible and immediately dropped, b) can be evaluated at a lower-level model (MP2 level), and c) can be evaluated at CC level. 
\item Find an optimal representation of the virtual space for each $ij$-pair, in which the derived tensors (amplitudes, residuals, or integrals) are dense.
\item Find \textblack{the} optimal factorization of terms in amplitude equations.
\item Perform only those tensor contractions \textblack{that} lead to non-zero results (including integral transformation and amplitude equations evaluation). 
\end{enumerate}

In the local pair natural orbital CCSD method, the occupied orbitals are localized (for example, by \textblack{the} Pipek--Mezey or Foster--Boys algorithm\textblack{s} \cite{foster-boys,localization_pipek})\textblack{,} and the virtual space is constructed specifically for a given occupied orbital pair. First, the virtual orbitals are transformed to a local basis of projected atomic orbitals (PAO) $|\tilde{\mu}\rangle$ as $|\tilde{\mu}\rangle = (1-\sum_i |i\rangle\langle i|)|\mu\rangle$, \cite{local_ccsd, local_ccsd2, local_ccsd3}, which are a priori local and can be easily
used to form domains corresponding to local occupied orbitals. In the next step\textblack{,} pair natural orbitals (PNO\textblack{s}) are constructed.\cite{pno1,pno2,dlpno-sparse,dlpno-sparse2,dlpno-nevpt2,werner-pno-lcc} \textblack{Therefore}, in Eqs. \ref{cc8a} and \ref{cc8b}, instead of virtual index $a,b,…$ we will get pair-specific PNOs {$\tilde{a}_{ij}$}, {$\tilde{b}_{ij}$}, … where we assume that the size of PNO space $N({\rm PNO}(ij)) \ll N({\rm MO_{\rm virt}})$. The PNO spaces are obtained from pair density matrices $\mathbf{D}^{ij}$
\begin{equation}
\mathbf{D}^{ij} = \mathbf{\tilde{T}}^{ij}\mathbf{T}^{ij\dagger} + \mathbf{\tilde{T}}^{ij\dagger}\mathbf{T}^{ij} \;,
\label{dlpno_dens}
\end{equation}
where $\mathbf{T}^{ij}$ are MP2 amplitudes in PAO basis
\begin{equation}
{T}^{ij}_{\mu\nu} = \frac{(i\tilde{\mu}|j\tilde{\nu})}{\epsilon_{\tilde{\mu}}+\epsilon_{\tilde{\nu}}-f_{ii}-f_{jj}} \;,
\label{dlpno_t2ampl}
\end{equation}
where $f_{ii}$ and $f_{jj}$ are occupied orbital energies, and $\epsilon_{\tilde{\mu}}$ or $\epsilon_{\tilde{\nu}}$ are PAO orbital energies obtained by diagonalization of the Fock matrix transformed to PAO space.
The transformation matrices $\mathbf{d}^{ij}$ transforming orbitals from PAO space to PNO space\textblack{,} and occupation numbers ${n}^{ij}$ correspond to eigenvectors and eigenvalues of the diagonalized matrix $\mathbf{D}^{ij}$\textblack{,}
\begin{equation}
\mathbf{D}^{ij}{\mathbf{d}}^{ij} = {n}^{ij}\mathbf{d}^{ij} \;.
\label{dlpno_diag}
\end{equation}
The introduction of $ij$-specific PNO spaces, which are mutually non-orthogonal, leads to more complicated expressions in the amplitude equations because we also need to involve overlap matrices between different PNO spaces. The matrix transforming $ij$-PNO space to $kl$-PNO space $S^{ij;kl}_{\tilde{a}_{ij}\tilde{b}_{kl}}$ is defined as
\begin{equation}
\mathbf{S}^{ij;kl} = {\mathbf{d}^{ij}}^{\dagger}\mathbf{S}^{\rm PAO}{\mathbf{d}^{kl}} \;,
\label{dlpno_transf}
\end{equation}
where $\mathbf{S}^{\rm PAO}$ is \textblack{the} PAO overlap matrix.
\textblack{Therefore}, when considering the PNO space, the term in Eq. \ref{cc9} 
can take the form
\begin{equation}
  \frac{1}{4} v^{\tilde{e}_{mn}\tilde{f}_{mn}}_{mn} t^{ij}_{\tilde{e}_{ij}\tilde{f}_{ij}} t^{mn}_{\tilde{a}_{mn}\tilde{b}_{mn}} S^{mn;ij}_{\tilde{a}_{mn}\tilde{a}_{ij}}
  S^{mn;ij}_{\tilde{b}_{mn}\tilde{b}_{ij}}
  S^{mn;ij}_{\tilde{e}_{mn}\tilde{e}_{ij}}
  S^{mn;ij}_{\tilde{f}_{mn}\tilde{f}_{ij}} \;.
  \label{dlpno_term1}
\end{equation}
However, it is also possible to transform \textblack{the} $e,f$ indices from $ij$- to $mn$-PNO space first and then to perform the contraction of the integral with the first amplitude.
The complex structure of these terms significantly expands the possibilities of how \textblack{they} can be factorized. The cost is also affected by the integral evaluation, which depends on the available memory. We can pre-compute not only $v^{\tilde{e}_{ij}\tilde{f}_{ij}}_{ij}$ type integrals, but also some mixed PNO space integrals (only those which \textblack{that} be needed). 

In our work, we assume that we will utilize a larger number of nodes, so we will have more memory and computing effort available. In that case, we can afford tighter thresholds, leading to larger domains and PNO spaces, which means higher precision in the recovery of the correlation energy. For the implementation of DLPNO formulations in TAMM, we employed a code generator implemented in Python that transforms the canonical equations by the transformation rules for various spaces (i.e., PNO, PAO, etc.). Using a code generator allowed us to systematically convert equations while automatically applying an operation cost minimization algorithm for single-term optimization. We anticipate that these kinds of code generators on the high-level equations \textblack{will} enable trying different transformations and automatically choosing the best performing one. While this implementation is in the early stages, we were able to validate the correctness of \textblack{the} generated code using our infrastructure to directly compare the results with \textblack{their} canonical counterparts. 

The perturbative correction (T) described in Sec. 4.1.2 is in the local version evaluated using the same equations with some differences.\cite{doi:10.1063/1.4821834} While converged T$_1$ and T$_2$ amplitudes are obtained in their PNO space, the virtual indices in terms $A^{abc}_{ijk}$ or $B^{abc}_{ijk}$ are represented in triples natural orbitals (TNO\textblack{s}), which are computed from triplet density \textblack{matrices} obtained from three pair density matrices\textblack{,} $\mathbf{D}^{ijk}=1/3 (\mathbf{D}^{ij}+\mathbf{D}^{ik}+\mathbf{D}^{jk})$. In order to perform contractions of integrals and amplitudes in Eqs. \ref{tensort} and \ref{tensort2}, we need to involve transformation matrices between PNO and TNO spaces $\mathbf{S}^{ij;klm}$, which are computed the same way as in Eq. \ref{dlpno_transf} where $\mathbf{d}^{kl}$ is \textblack{replaced} by $\mathbf{d}^{klm}$, transforming PAO orbitals to TNO space. Similar to \textblack{the} DLPNO CCSD implementation, we leveraged the TAMM framework's dense tensor infrastructure to represent the perturbation correction implementation with block-sparse computation. Using the PNO representation of the amplitudes from \textblack{the} CCSD implementation, we implemented the DLPNO formulation of the canonical equations.

\finalresponse{For the development of DLPNO-based methods {\color{black}in NWChemEx}, we utilized specialized view tensor construction to avoid redundant storage of the transformation tensors, especially for the transformation required for the occupied pair indices. Figure \ref{fig:view_tensor_const} shows an example use case specific to DLPNO CCSD equations. In this example, we are using a view tensor over the large \code{Sijkl} transformation tensors. Depending on the pair indices used within the equations, various kinds of transformations are required. In this example, the constructed \code{Sijki} tensor (line 17)  has values from \code{Sijkl} if the first index of the first pair matches the second index of the second pair. If not, the corresponding indices are set to zero. In this case, there was no need to translate the blocks or update the tensor since the tensor was used as read-only. However, users can provide special constraints on both the translation of the blocks \textblack{and access} to these blocks. By using this feature, we were able to reduce the memory footprint of DLPNO methods and add additional constraints to leverage sparsity within the computation. Since TAMM allows mixed usage of specialized tensors with other tensor types in all supported operations, the representation and execution patterns of the DLPNO CCSD equations in TAMM did not have to change in such scenarios.} 

\begin{figure}[t]
\centering
\begin{lstlisting}
// Translate lambda for translating BlockId to reference tensor
// No translation is needed for this case 
auto no_translate = [](IndexVector blockid) -> IndexVector {
  return blockid;
};

// Lambda for specifying how to put values to the reference tensor
// For read-only usage leave it empty
auto put_copy_func = [](const BlockSpan<T>& in, BlockSpan<T>& out, const IndexVector& blockid) -> void {};

// Lambda for specifying how to get values from the reference tensor
auto get_S_is_si = [=](const BlockSpan<T>& from_span, BlockSpan<T>& to_span, const IndexVector& blockid) -> void{
    // Only copy the values where first index of first pair
    // and last index of second pair matches
};

Tensor<T> Sijki{Sijkl, {is, si, a_is, a_si}, no_translate, get_S_is_si, put_copy_func};
\end{lstlisting}
\caption{View \textblack{the tensor} construction for Sijkl transformation tensors in DLPNO CCSD equations}
\label{fig:view_tensor_const}
\end{figure}

\subsubsection{Time-dependent \textblack{c}oupled-\textblack{c}luster \textblack{m}ethod}

In addition to the stationary and frequency-dependent formulations of CC theory, one could witness significant progress on our end in developing a time-dependent CC method for simulating real-time dynamics.

The TDCC method has been studied in the context of the  CC linear-response theory,\cite{monkhorst77_421,jorgensen90_3333, mukherjee1979response, dalgaard1983some, koch1990coupled} 
\textblack{x}-ray spectroscopy and Green's function theory,\cite{schonhammer1978time,nascimento2017simulation, rehr2020equation, vila2021equation, vila2022real} nuclear physics,\cite{hoodbhoy1978time,hoodbhoy1979time,pigg2012time}  condensed matter physics,\cite{arponen83_311}  and quantum dynamics of 
molecular systems in external fields.\cite{huber2011explicitly,kvaal2012ab,pedersen2019symplectic,kristiansen2020numerical,sato2018communication, pathak2020time}
These studies have also initiated an intensive effort toward understanding many aspects of the TDCC formalism, including addressing fundamental problems such as the form of the action functional, the form of the time-dependent molecular orbital basis, \textblack{the} physical interpretation of time-dependent orbitals, various-rank approximations of the cluster operator, and \textblack{the} numerical stability of time integration algorithms. One of the milestone achievements in developing \textblack{a} time-dependent CC formalism was  Arponen's action functional for the bi-variational coupled cluster formalism\cite{arponen83_311} and \textblack{the} following analysis by Kvaal considering time-varying orbitals.\cite{kvaal2012ab}

The time-dependent Sch{\"o}dinger equation (TDSE),
\begin{eqnarray}
&i {\partial_\tau}|\Psi(\tau)\rangle= H |\Psi(\tau)\rangle,
\label{tdse}
\end{eqnarray}
is an initial value problem,
which has a formal solution 
$|\Psi(\tau)\rangle = e^{-iH\tau}|\Psi\rangle,$    
with the initial condition, 
$|\Psi(0)\rangle= |\Psi\rangle$.
An efficient way of finding approximate solutions of the TDSE for many-electron systems is associated with the utilization of the time-dependent CC (TD-CC) Ansatz
\begin{equation}
|\Psi(\tau)\rangle =e^{T(\tau)}|\Phi\rangle \;.
\label{xxx1}
\end{equation}
Upon plugging TD-CC Ansatz into Eqn~\ref{tdse}, and projecting onto the excited determinants, one obtains 
\begin{eqnarray}
i {\partial_\tau} \langle\Phi_{i_1\ldots i_k}^{a_1 \ldots a_k}| T(\tau)|\Phi\rangle &=& \langle\Phi_{i_1\ldots i_k}^{a_1 \ldots a_k}|\overline {H}(\tau)|\Phi\rangle
\label{time_derivative}
\end{eqnarray}
where $T(\tau)$ is \textblack{a} time-dependent cluster operator,  $|\Phi\rangle$ is a time-independent reference wave function in our consideration, and 
$\overline{H}(\tau)=e^{-T(\tau)} H e^{T(\tau)}$, is the similarity transformed time-dependent Hamiltonian, which is the generator for
the time evolution.

The similarity transformed CC Hamiltonian is non-Hermitian. Therefore, the computation 
of observables in this framework requires a bi-variational approach,\cite{arponen83_311,stanton93_7029} i.e., both the bra and ket states must be varied independently.
The expectation value of any observable in \textblack{the} TDCC framework is defined as
\begin{eqnarray}
\langle O\rangle=\left\langle \Phi|\left (1+\Lambda(\tau)\right) \overline{O}(\tau) |\Phi \right\rangle    
\end{eqnarray}
where $\Lambda$ is a de-excitation operator.
For obtaining equations for the $\Lambda$, we write down the TDSE,
$-i {\partial_\tau}\langle \Psi^\prime(\tau)|= \langle \Psi^\prime(\tau)| H,$
for the bra vector, with $H= H^\dag$,
\begin{eqnarray}
&-i {\partial_\tau}\langle \Phi|(1+\Lambda(\tau)) e^{-T(\tau)}= \langle \Phi|(1+\Lambda(\tau)) e^{-T(\tau)} H
\end{eqnarray}
Multiplying by $e^{T(\tau)}$ from the right, and right projecting to the excited determinants obtains
\begin{eqnarray}
-i{\partial_\tau}\langle\Phi| \Lambda(\tau) |\Phi^{i_1\ldots i_k}_{a_1 \ldots a_k}\rangle
&=& \langle\Phi|(1+\Lambda(\tau))\overline {H}(\tau)|\Phi^{i_1\ldots i_k}_{a_1 \ldots a_k}\rangle
\label{time_derivative2}
\end{eqnarray}

\finalresponse{The right-hand side of Eqn. \ref{time_derivative}, is the same as the stationary CC amplitude equations and hence implemented using TAMM\textblack{,} similar to how it is implemented for the stationary theory as shown in Sec.~\ref{ccsd_section}.
However, the time-dependent CC amplitudes are naturally complex
since the underlying equation we are solving is the TDSE.
Furthermore, one-body and two-body Hamiltonian matrix elements are real-valued quantities. 
TAMM support for mixed complex-real tensor operations helps avoid the explicit handling of the real 
and imaginary data separately.}
For example, a tensor contraction for the TDCC method\textblack{,}
\begin{eqnarray}
(R(a,\, b,\, i,\, j\,)   &+=& 0.5*v(a,\, b,\, c,\, d\,)*t(c,\, d,\, i,\, j\,))    
\end{eqnarray}
\finalresponse{can be expressed exactly the same way as in the stationary CC theory as shown in Eqn~\ref{tamm_cc11}.
However, in this case, R and t are complex-valued, while v is a real-valued quantity.}

\finalresponse{We have propagated the time-dependent CC amplitudes using a \textblack{first}-order Adams--Moulton method, which is an implicit time-propagator. This algorithm requires copying and swapping of the real and complex parts of time-dependent amplitudes in each micro-iterations. TAMM does not directly support swapping the real and imaginary parts of each element of a complex tensor. However, in this case, we used the $block\_for$ construct of TAMM to perform the swap in parallel. A \textblack{specialized} C++ lambda expression, as shown in Figure \ref{fig:rteom_swap}, is provided to the $block\_for$\textblack{,} which simply specifies that the real and imaginary parts of each element in a given block need to be swapped and copied to (or update\textblack{d}) the corresponding element in the destination tensor block.}

\begin{figure}[t]
\centering
\begin{lstlisting}
    // src, dest tensors are given
    auto complex_swap_lambda = [&](const IndexVector& blockid) {

      // assert TensorType is complex
      const auto dsize = dtensor.block_size(blockid);
      span<TensorType> sbuf{stensor.access_local_buf(), dsize};
      span<TensorType> dbuf{dtensor.access_local_buf(), dsize};

      for(auto i = 0; i < dsize; i++) {
        TensorType val(sbuf[i].imag(), sbuf[i].real());
        if(update) dbuf[i] += val;
        else dbuf[i] = val;
      }
    };

    block_for(execution_context, dtensor(), complex_swap_lambda);

\end{lstlisting}
\caption{Parallel element-wise swap and copy operation between complex tensors}
\label{fig:rteom_swap}
\end{figure}


\finalresponse{The TDCC method performs time propagation for a sufficiently long duration to obtain well-resolved spectra.
However, obtaining computing time for such an extended period in a single run is unlikely when using shared computing resources. In addition, it is highly beneficial to track simulations to adjust and optimize various parameters in between different runs of the same simulation. The ability to checkpoint and restart simulations\textblack{,} hence\textblack{,} plays a significant role in \textblack{the} efficient utilization of computing resources. The TAMM library provides parallel file I/O operations on tensors, which helps with checkpointing and restarting the TDCC calculations. We simply use the TAMM interface to write and read tensor data to and from \textblack{the} disk\textblack{,} in as shown in Fig.~\ref{fig:cc_tensor_io}. The underlying TAMM infrastructure handles the \textblack{parallelization} aspects of these I/O operations.}


\begin{figure}[t]
\centering
\begin{lstlisting}
    tamm::write_to_disk (tensor, tensor_filename);
    tamm::read_from_disk(tensor, tensor_filename);
\end{lstlisting}
\caption{\textblack{P}arallel file I/O operations in TAMM\textblack{.}}
\label{fig:cc_tensor_io}
\end{figure}

The TAMM infrastructure has been utilized to implement the time evolution of the $T(\tau)$ operator. We have considered singles (S) and doubles (D) excitation approximations in our implementation. Our implementation uses Cholesky-decomposed two-body electron repulsion matrix elements,\cite{bo1, bo2} and we exploit spin-explicit formalism to evaluate tensor contractions of matrix elements of various operators.

\subsubsection{Coupled \textblack{c}luster Green's \textblack{f}unction}

The correlated formulations of Green’s function methods are indispensable elements of the computational infrastructure needed not only to calculate ionization potentials, electron affinities, and excitation energies but also as quantum solvers for various embedding formulations. The CC formalism provides a natural platform for the development of the one-body Green’s function and the introduction of high-rank correlation effects. \cite{nooijen92_55,nooijen93_15,nooijen95_1681,meissner93_67}
Without loss of generality, here we will focus only on the retarded part of Green’s function (the advanced part can be developed in an analogous way) defined by matrix elements $G^R_{pq}(\omega)$:
\begin{equation}
G^R_{pq}(\omega) =
\langle \Psi_g | a_q^\dagger (\omega + ( H - E_0 ) - i \eta)^{-1} a_p | \Psi_g \rangle \;,
\label{gf0}
\end{equation}
where $E_0$ is the corresponding ground-state energy for \textblack{the} $N$-electron system, $\eta$ is a broadening factor, and $|\Psi_g\rangle$ represents \textblack{the} ground-state of \textblack{the} $N$-electron system. In CC we are using different parametrizations for the bra ($\langle\Psi_g|$) and ket 
($|\Psi_g\rangle$) wave functions,
\cite{arponen83_311,stanton93_7029}
i.e., 
\begin{eqnarray}
\langle \Psi_g | &=& \langle \Phi | (1+\Lambda)e^{-T}  \label{biv1} \\
| \Psi_g \rangle &=& e^T | \Phi \rangle,
\label{biv2}
\end{eqnarray}
which leads to the following form of retarded part of the CC Green’s function (CCGF) 
\cite{nooijen92_55,nooijen93_15,nooijen95_1681,meissner93_67,kkgfcc1,kkgfcc3}
\begin{eqnarray}
G^R_{pq}(\omega) = 
\langle\Phi|(1+\Lambda) \overline{a_q^{\dagger}} (\omega+\bar{H}_N- \text{i} \eta)^{-1} 
	\overline{a_p} |\Phi\rangle. 
\label{gf1}
\end{eqnarray}
The similarity transformed operators here $\overline{A}$ ($A = H, a_p, a_q^{\dagger}$) are defined as $\overline{A} = e^{-T} A ~e^{T}$ (the $\overline{H}_N$ stands for a normal product form of $\overline{H}$). The numerically feasible algorithms for calculating (\ref{gf1})
employ $\omega$-dependent auxiliary  operators $X_p(\omega)$
\begin{eqnarray}
X_p(\omega) 
&=& X_{p,1}(\omega)+X_{p,2}(\omega) + \ldots \notag \\
&=& \sum_{i} x^i(\omega)_p  a_i  + \sum_{i<j,a} x^{ij}_a(\omega)_p a_a^{\dagger} a_j a_i +\ldots ,
\;\; \forall_p \label{xp} 
\end{eqnarray}
that satisfy equations 
\begin{eqnarray}
(\omega+\overline{H}_N - \text{i} \eta )X_p(\omega)|\Phi\rangle = 
	\overline{a_p} |\Phi\rangle.  \label{eq:xplin} 
\end{eqnarray}
Using  these operators matrix elements 
can be expressed in a simple form 
\begin{eqnarray}
G^R_{pq}(\omega) = 
\langle\Phi_g|(1+\Lambda) \overline{a_q^{\dagger}} X_p(\omega) |\Phi_g\rangle.
\label{gfxn2}
\end{eqnarray}
In our implementation of CCGF formalism in the SPEC project, we approximate 
$\Lambda$ operator by $T^{\dagger}$.
The main numerical effort associated with constructing \textblack{a} retarded CC Green's function is associated with the need \textblack{to solve} a large number of independent linear equations, which in turn can contribute to efficient parallel schemes utilizing multiple levels of parallelism. 
 \finalresponse{Our CCGF algorithm consists of an outer loop that checks for the convergence of the entire calculation. We refer to iterations of this loop as levels. Each level mainly consists of two loops which are the most computationally intensive part of the entire calculation. The first loop goes over the frequencies ($\omega’s$) sampled using the adaptive midpoint refinement strategy.~\cite{mor_spectrum2017} The second loop goes over all the orbitals ($p’s$) and the CCGF singles and doubles equations are solved for all ($\omega,\,p$) pairs for a given level in this loop. Since the bulk of the computation here is associated with high\textblack{-}dimensional tensor contractions, we express the tensor operations similar to the cases in Fig.~\ref{fig:tensor_exec} and Eqn.~\ref{tamm_cc11}\textblack{,} even though the tensor operations in CCGF involve a mix of complex and real data types. This demonstrates the uniform operation representation provided by TAMM for operations on mixed data types.}

 \finalresponse{We refer to the list of all ($\omega,\,p$) pairs in a given level as CCGF tasks each of which can be executed independently. Each task solves the CCGF singles and doubles equation\textblack{s} until convergence for a given ($\omega,\,p$) pair. All the $p$ orbitals for a given $\omega$ are divided across process groups that are set up at the beginning of each level. Currently, the user is responsible for explicitly creating MPI process groups and using them to construct the TAMM \boldit{process group} and \boldit{execution context} objects which are in turn used for executing each task. Each TAMM \boldit{process group} now contains a subset of the resources (processes) provided to the CCGF calculation. In our CCGF implementation, the size of a process group for computing each task is determined automatically for a given problem size (i.e., number of occupied \textblack{and} virtual orbitals) and the resources (total number of nodes, processes, \textblack{and} GPUs per node) provided for that run. All process groups are created to be the same size. We are actively working on providing abstractions in TAMM for expressing computation to be divided across process groups without the user having to create them manually using MPI. Once the \boldit{execution context} objects representing different process groups are created, all CCGF tasks are distributed across \textblack{the} available process groups and executed using different execution contexts.}
 
 \finalresponse{The resulting tensors from each task are stored on disk using the parallel file I/O routines provided by TAMM for writing and reading distributed tensors to and from \textblack{the} disk\textblack{,} as shown in Figure \ref{fig:cc_tensor_io}. This enables restarting the CCGF calculation at any point in the calculation. If the set of tasks in a given level \textblack{is} not completed, the subsequent CCGF run upon restart will create process groups only for the remaining tasks and execute them. Another feature provided by TAMM used in CCGF is batched parallel file I/O operations. Several tensors for each task are written to disk as part of the CCGF calculation. The number of tensors written to disk grows with increasing CCGF problem size and the number of frequencies ($\omega$) to be solved. Even though each tensor is read/written using parallel file I/O, the operation uses a small subset of available nodes/processes to perform the read/write operation for a single tensor. This is because the entire set of available resources used for the CCGF calculation \textblack{is} not required to read/write even large tensors from/to disk. This leads to resource under-utilization since most of the resources are idle while a subset of them are performing the parallel file I/O operation on one distributed tensor at a time. To address this issue, TAMM provides high-level batched tensor I/O routines that can read several hundred tensors from disk concurrently by automatically dividing the total available resources into smaller process groups. Since all tensors that need to be read/written are not necessarily of the same size, variable-sized process groups are dynamically created based on the sizes of each tensor in a given list of tensors. Each of these process groups reads/writes a tensor using parallel file I/O within the group. Several tensors are read/written concurrently by different process groups, leading to significantly better resource utilization and overall improvement in the total time spent in file IO. The user would express such an operation in TAMM as shown in Figure \ref{fig:batched_tensor_io}. The operation is expressed exactly the same way as one would express it for reading or writing a single tensor (e.g., Figure \ref{fig:cc_tensor_io}), with the only obvious difference being that a list of tensor handles and corresponding filenames need to be provided in this case.}
\begin{figure}[t]
\centering
\begin{lstlisting}
    tamm::write_to_disk (tensor_list, tensor_filenames_list);
    tamm::read_from_disk(tensor_list, tensor_filenames_list);
\end{lstlisting}
\caption{Batched parallel file I/O operations in TAMM}
\label{fig:batched_tensor_io}
\end{figure}
\finalresponse{Reference~\citenum{GFCCLib21} describes the highly scalable CCGF implementation developed using TAMM in detail\textblack{,} along with the performance and scalability analysis on OLCF Summit. The implementation is also publicly available.~\cite{gfcc_github}}

\finalresponse{This section provides an overview of some important CC methods and discusses key aspects of their implementations using TAMM. While frameworks such as TiledArray, CTF, DISTAL,\textblack{and} ExaTensor provide similar features in how tensor operations in CC methods can be represented,  TAMM differs from them as explained in Section \ref{sec:feature_compare_other_tensor_libs}. To \textblack{the} best of our knowledge, these are the first complete scalable distributed implementations of the discussed CC methods that can be run on different GPU architectures while also using different parallel programming models.}


\section{\textblack{PERFORMANCE COMPARISON WITH OTHER TENSOR ALGEBRA FRAMEWORKS}}
\label{sec:performance_comparison}

\finalresponse{We have provided feature comparisons with other distributed tensor algebra frameworks in Section \ref{sec:feature_compare_other_tensor_libs}, where key library features such as tensor-construction and tensor-operation specification, hardware support, and the underlying parallel programming model backends (for distributed data management) are compared. This section details the performance comparison results with other distributed tensor algebra frameworks, 
namely, TiledArray (TA) and Cyclops Tensor framework (CTF). We did not consider comparisons with ExaTensor since the development page~\cite{exatensor_repo} states that the library has pending performance issues as well as numerous problems with existing MPI-3 implementations at the time of this writing.}

\finalresponse{All our experiments were performed on the National Energy Research Scientific Computing Center (NERSC) Perlmutter supercomputer. The GPU partition was used for all the experiments. Each node in this partition has a 64-core AMD EPYC 7763 CPU, 256 GB of DDR4 DRAM, and 4 Nvidia A100 GPUs, each with 40 GB of HBM2e RAM. Each GPU node is connected to \textblack{four} NICs, and the GPU nodes are connected using the Slingshot 11 interconnect. The TAMM experiments were configured to run with \textblack{four} MPI ranks per node \textblack{and} \textblack{one} MPI rank per GPU. The TiledArray experiments were configured with \textblack{four} MPI ranks per node (\textblack{one} MPI rank per GPU) and \textblack{two} threads per rank since that is the TiledArray developer\textblack{'s} recommended configuration for Perlmutter. The CTF implementation provided the best performance when using either 8 or 12 MPI ranks per node (2 or 3 MPI ranks per GPU), depending on the problem size and the number of nodes used in the experiment. For a given dimension size (N) and node count, we ran CTF using both 8 and 12 MPI ranks per node and chose the best timing out of the two. All codes were compiled with GCC 11.2, CUDA 11.7, and cray-mpich 8.1.}

\finalresponse{We compared the performance of a tensor contraction shown in Equation~\ref{4d_tc_bm}, which is one of the most expensive operations in many CC calculations. For benchmarking purposes, each dimension of the 4D tensors in Eqn.~\ref{4d_tc_bm} is chosen to be of the same size (N) resulting in a total size of $(N^4)$ double-precision floating- point elements for each tensor. The input tensors are filled with random data in each case. We used the readily available TA and CTF codes~\cite{tiled_array,ctf_github} for benchmarking equation~\ref{4d_tc_bm} and ensured that the ordering of the indices used in the contraction is consistent across the different \textblack{benchmark} codes. The TA benchmark allows specifying the number of blocks for each dimension. For a given dimension size (N) and node count, we ran the TA benchmark several times using varying number of blocks along each dimension in each run and chose the timings for the best performing case for comparing against TAMM.}

\begin{equation}
 C(a,b,c,d) += A(a,b,m,n) * B(m,n,c,d) 
\label{4d_tc_bm}
\end{equation}


\begin{figure}[!ht]
\centering
\begin{tabular}{@{}c@{}}
\includegraphics[width=0.47\textwidth]{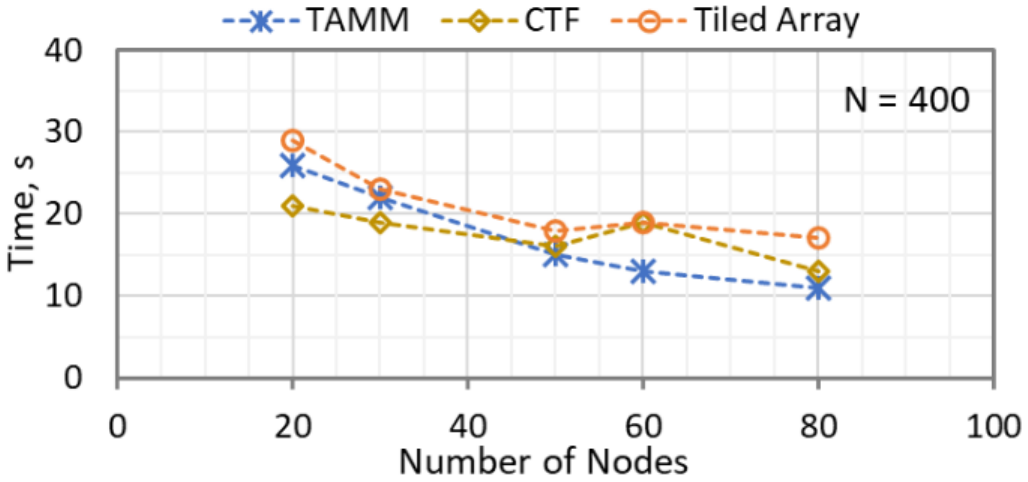} \\[\abovecaptionskip]
\small(a)
\label{fig:tamm_scale_4d_tensor_perlmutter_a}
\end{tabular}
\hfill
\begin{tabular}{@{}c@{}}
\includegraphics[width=0.47\textwidth]{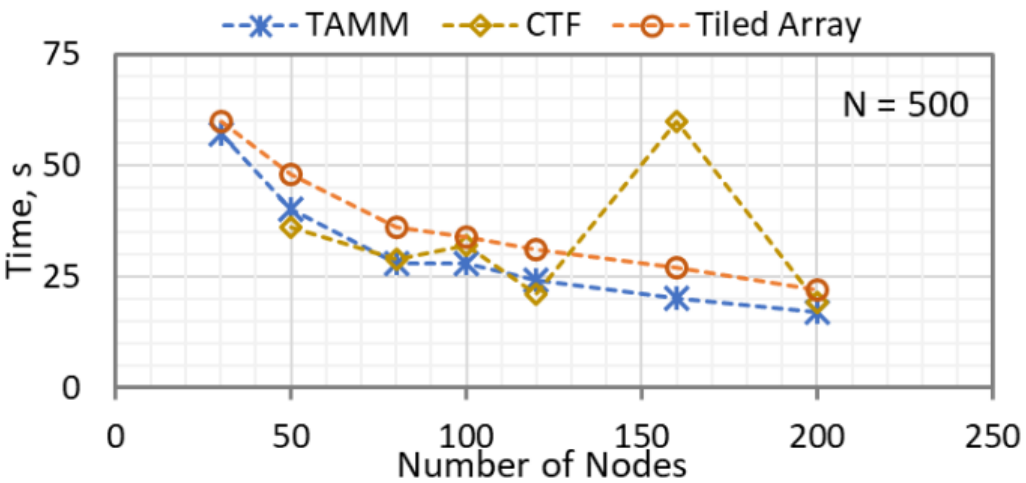} \\[\abovecaptionskip]
\small(b)
\end{tabular}
\hfill
\begin{tabular}{@{}c@{}}
\includegraphics[width=0.47\textwidth]{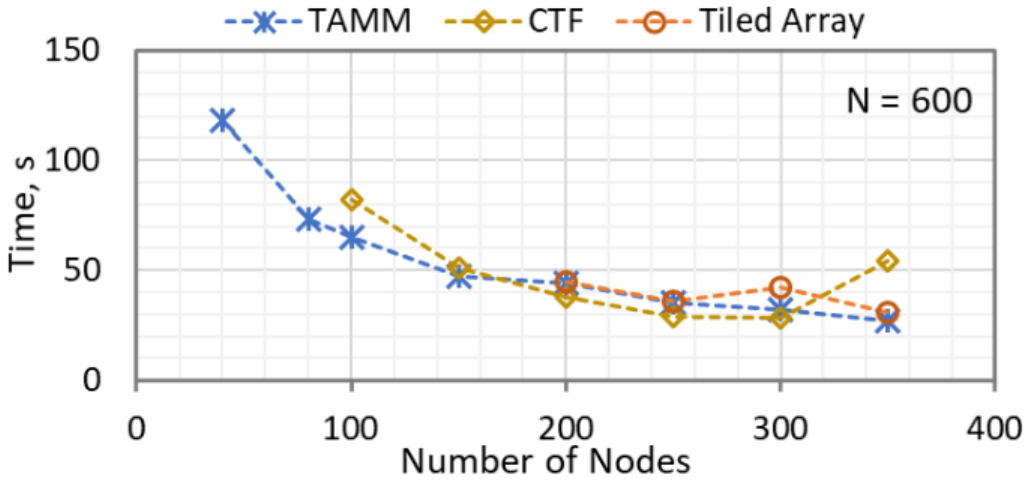} \\[\abovecaptionskip]
\small(c)
\end{tabular}
\hfill
\begin{tabular}{@{}c@{}}
\includegraphics[width=0.47\textwidth]{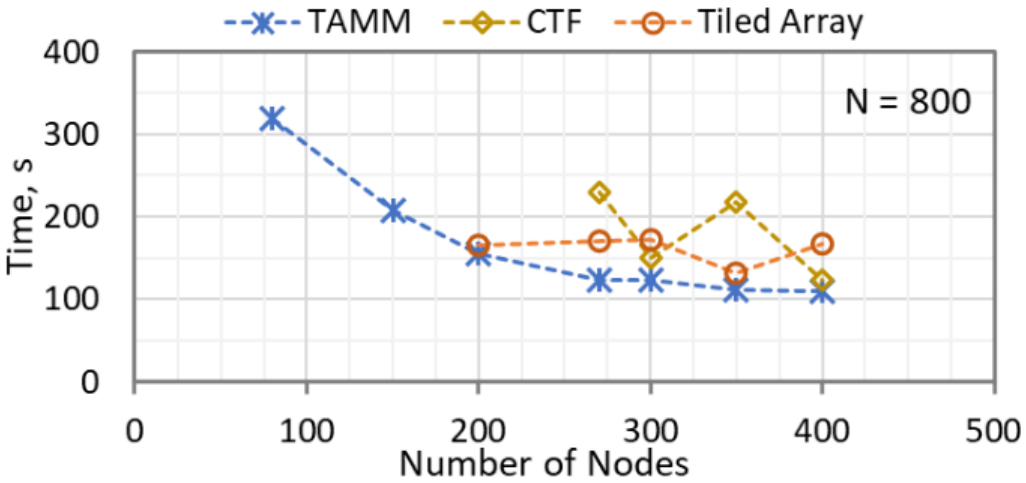} \\[\abovecaptionskip]
\small(d)
\end{tabular}
\hfill
\caption{Scaling of 4D tensor contraction with TAMM and comparison with two other implementations.}  \label{fig:tamm_scale_4d_tensor_perlmutter}
\end{figure} 

The performance results for TAMM, TiledArray, and CTF implementations of equation~\ref{4d_tc_bm} are shown in Figure \ref{fig:tamm_scale_4d_tensor_perlmutter}. Plots for $N=$ 400, 500, 600, and 800 are also shown on different Perlmutter node counts. 
 With \textblack{an} increasing number of nodes in order to strong scale for a given problem size ($N$), TAMM demonstrated consistent performance improvements. In contrast, TA and CTF exhibited some oscillating behavior. The TAMM implementation strong scales consistently for a given problem size ($N$), and provides competitive performance compared to TA and CTF\textblack{,} especially for large problem sizes and node counts.

\finalresponse{Table ~\ref{Table:min_node_count} shows the minimum number of  nodes required on Perlmutter to run each of the three implementations for the tensor contraction shown in (\ref{4d_tc_bm}) for large problem sizes ($N=600$  and $N=800$).  As $N$ increased, the required minimum number for running the benchmark in the case\textblack{s} of TA and CTF was much higher than expected in comparison to TAMM.
At the time of this writing, CTF crashed due to segmentation faults on node counts less than 100 for $N=600$ and node counts less than 270 for $N=800$. Similarly, TA also required a minimum of 40 nodes for $N=600$ and 200 nodes for $N=800$\textblack{,} whereas TAMM required only 25 and 80 nodes\textblack{,} respectively\textblack{,} for those cases.}

\begin{table}[!ht]
    \centering
        \caption{Minimum number of Perlmutter nodes required to run the dense 4D tensor contraction shown in Equation \ref{4d_tc_bm}.}
    \begin{tabular}{llll}
    \hline
    \hline
        & \multicolumn{3} {l}{\textbf{Number of Nodes}} \\
     \hline   
        \textbf{Dim size (N)} & \textbf{TAMM} & \textbf{TA} & \textbf{CTF} \\
 \hline       
        600 & 25 & 40 & 100  \\
        800 & 80 & 200 & 270 \\
        \hline
        \hline
    \end{tabular}

    \label{Table:min_node_count}
\end{table}

\finalresponse{We also compare the performance of the contraction in Equation~\ref{4d_tc_bm} with COSMA~\cite{cosma} by representing the contraction as a distributed matrix-multiply operation. COSMA is a state-of-the-art distributed matrix-multiply library \textblack{that} is capable of running on GPUs. Equation~\ref{mm_bm} shows the equivalent of Equation~\ref{4d_tc_bm} represented as a distributed matrix-multiply operation in COSMA\textblack{,}}

\begin{equation}
C(i,j) +=  A(i,k) * B(k,j) 
\label{mm_bm}
\end{equation}

We use the COSMA provided distributed matrix-multiply miniapp.~\cite{cosma_github} The size of each dimension for \textblack{the} COSMA benchmark was chosen as $N^2$ in order to match the overall sizes for each tensor (i.e. $(N^4)$) in Equation~\ref{4d_tc_bm}. Table ~\ref{Table:tamm_cosma} shows the values for $N$ considered in the comparison. At the time of this writing, we observe that COSMA, like the TiledArray and CTF implementations, also \textblack{requires} a certain minimum number of nodes\textblack{,} which is significantly higher than what is required by the corresponding TAMM implementation for a given problem size (N). We observe that the TAMM performance is nearly identical to COSMA for $N=400$ and about 10\% better than COSMA for $N=500$. For larger problem sizes $N=600$ and $N=800$, \textblack{the} COSMA implementation ran out of memory on 300 and 400 nodes, respectively. The DISTAL~\cite{DISTAL22} work also provides a performance comparison with COSMA. The DISTAL authors mention that COSMA generally achieved about 10\% higher performance in comparison to DISTAL using up to 256 nodes of the Lassen supercomputer. Since the code for DISTAL is not in the public domain at the time of this writing, a direct performance comparison with DISTAL is currently not possible.

\begin{table}[!ht]
    \centering
        \caption{Performance comparison with COSMA.}
    \begin{tabular}{llll}
    \hline
    \hline
        \textbf{Dim size (N)} & \textbf{Node Count} & \textbf{TAMM} & \textbf{COSMA} \\ \hline
        300 & 50 & 6 & 5 \\ 
        400 & 100 & 9 & 9 \\ 
        500 & 100 & 28 & OOM* \\ 
        500 & 200 & 17 & 19 \\ 
        600 & 300 & 33 & OOM* \\ 
        600 & 400 & 28 & OOM* \\ 
 \hline
 \hline
    \end{tabular} \\
    \footnotesize{* Did not run to completion due to \textblack{out-of-memory} errors}

    \label{Table:tamm_cosma}    
\end{table}

{\color{black}
\textbf{CCSD benchmark:} We implemented our Cholesky-CCSD equations as tensor expressions in TA and CTF to further investigate the performance on a real application. We ran calculations of Ubiquitin-DGRTL~\cite{ubiquitin1231,ubi_dgrtl} molecule with the aug-cc-pVDZ basis (O=146, V=1096, 7810 Cholesky vectors) on up to 350 nodes of Perlmutter using the best performance parameters for TA and CTF. With TAMM, for a single CCSD iteration on 240 nodes, we observed an 80\% speed-up over CTF while TA ran out of memory. On 350 nodes, we observed nearly identical timing with CTF and 2.5x speed-up over TA.}

\section{\textblack{CONCLUSION}}
\label{sec:conclusion}

We have introduced and discussed the Tensor Algebra for Many-body Methods 
framework that enables scalable and performance-portable implementations of important computational chemistry methods on modern large-scale heterogeneous high-performance computing systems. We described the TAMM framework in detail by introducing a tensor algebra interface that provides \textblack{a} high-level representation of the tensor algebra operations as an embedded domain-specific language. This interface \textblack{enables the} separation of concerns between scientific application development and high-performance computing development efforts. The domain scientist or scientific application developer can focus on the method development instead of the performance optimization details, whereas the HPC developers focus on the underlying algorithms and optimizations. Later, we presented our modular infrastructure that allows the implementation of different optimizations on tensor data distribution, execution, and scheduling of tensor operations for efficient execution on modern heterogeneous HPC platforms. We also compared the features of TAMM against several other distributed tensor algebra frameworks. Through various case studies, we showcased the performance and productivity benefits of using the TAMM framework for implementing complete ground- and excited-state electronic structure methods that are expressed as operations on tensors. \finalresponse{Finally, we evaluated the performance of TAMM against other distributed tensor algebra frameworks and demonstrated it's competitiveness and scalability on large problem sizes and node counts on NERSC Perlmutter.}

\section*{\textblack{ACKNOWLEDGMENTS}} 

This research was supported by the Exascale Computing Project (\textblack{No.} 17-SC-20-SC), a collaborative effort of the U.S. Department of Energy Office of Science and the National Nuclear Security Administration, and by the Center for Scalable Predictive Methods for Excitations and Correlated Phenomena (SPEC), which is funded by the U.S. Department of Energy (DoE), Office of Science, Office of Basic Energy Sciences, Division of Chemical Sciences, Geosciences and Biosciences as part of the Computational Chemical Sciences (CCS) program at Pacific Northwest National Laboratory (PNNL) under \textblack{Grant No.} FWP 70942. PNNL is a multi-program national laboratory operated by Battelle Memorial Institute for the U.S. DoE under Contract \textblack{No.} DE-AC06-76RLO-1830. \textblack{N.P.B.} also acknowledges support from the Laboratory Directed Research and Development (LDRD) Program at PNNL.
%
%
We gratefully acknowledge the computing resources provided and operated by the Joint Laboratory for System Evaluation (JLSE) at Argonne National Laboratory, \textblack{the} Argonne Leadership Computing Facility, which is a DOE Office of Science User Facility supported under Contract \textblack{No.} DE-AC02-06CH11357, \textblack{and the} Oak Ridge Leadership Computing Facility, which is a DOE Office of Science User Facility supported under Contract \textblack{No.} DE-AC05-00OR22725.

\section*{\textblack{AUTHOR DECLARATIONS}}
\subsection{Conflict of Interest} 
The authors have no conflict of interest to declare.

\section{Data Availability}
The data that support the findings of this study are available from the corresponding authors upon reasonable request.

\bibliographystyle{h-physrev}
\bibliography{ref.bib}

\end{document}